K. Lodders

**Solar system abundances of the elements.**

In: Principles and Perspectives in Cosmochemistry.

Lecture Notes of the Kodai School on 'Synthesis of Elements in Stars' held at Kodaikanal Observatory, India, April 29 - May 13, 2008 (Aruna Goswami and B. Eswar Reddy eds.) Astrophysics and Space Science Proceedings, Springer-Verlag Berlin Heidelberg, 2010, p. 379-417 (ISBN 978-3-642-10351-3), 2010

# Solar System Abundances of the Elements

Katharina Lodders

Planetary Chemistry Laboratory, Dept. of Earth & Planetary Sciences and McDonnell Center for the Space Sciences, Washington University, Campus Box 1169, One Brookings Drive, Saint Louis, MO 63130, USA. lodders@wustl.edu

**Summary.** Representative abundances of the chemical elements for use as a solar abundance standard in astronomical and planetary studies are summarized. Updated abundance tables for solar system abundances based on meteorites and photospheric measurements are presented.

## 1 Motivations to Study Solar System Elemental Abundances

The investigations of what chemical elements exist in nature and in which quantities have a long history. The determination of elemental abundances in various celestial objects is still a very active field in astronomy, planetary science, and meteoritics. There are multiple motivations for studying the solar system abundances of the chemical elements. One reason to study this overall composition of the solar system is to understand how the diversity of planetary compositions, including that of our home planet, can be explained, since all planets in the solar system share a common origin from the material of the protosolar disk (the solar nebula).

The composition of the sun determines how the sun works and evolves over time, as composition influences the interior structure of the sun. Although the Sun is mainly composed of H and He, other heavy elements such as C, N, O, Ne, Fe, etc., are important opacity sources that influence the energy transport out of the sun through radiation and convection. The sun is a typical main sequence dwarf star and its composition is a useful baseline for comparison to abundances in other dwarf stars and to changes that appear in advanced stages of stellar evolution. For example, relative to the sun's composition, red giant stars show observable abundance variations that are the result of nucleosynthesis operating in giant stars, and these products have been dredged up from stellar interiors to the stellar exteriors.

The solar system abundances are a useful local galactic abundance standard because many nearby dwarf stars are similar in composition; however, in detail there are some stochastic abundance variations (e.g., Edvardsson et al. 1993, Nordstrom et al. 2004, Reddy et al. 2003, 2006). The term "cosmic" abundances should be avoided because abundances generally decrease with galactocentric distance. There are also abundance differences between our galaxy and galaxies at high red-shift; hence there is no generic "cosmic" composition that applies to all cosmic systems.

Finally, solar abundances are a critical test of nucleosynthesis models and models of Galactic chemical evolution. Ideally, such models should quantitatively explain the elemental and nuclide distributions of solar system matter.

The sun has the most mass (>99%) of the solar system objects and therefore it is the prime target for studying solar system abundances. Most elements can be measured in the sun's photosphere, but data from the solar chromosphere and corona, solar energetic particles, solar wind, and solar cosmic rays (from solar flares), help to evaluate abundances of elements that have weak absorption lines (because these elements are low in abundance or only have blended absorption lines in the photospheric spectrum).

Below we will see that meteorites, smaller rocks from asteroidal objects delivered to Earth, provide important information for solar system abundances of non-volatile elements. Other sources to refine solar system abundances are analysis of other solar system objects such as the gas-giant planets, comets and the interplanetary dust particles from comets. Outside the solar system, the compositions of hot B stars, planetary nebulae, Galactic cosmic rays (GCR), the nearby interstellar medium (ISM) and HII regions have been employed to amend the solar system abundances of some elements.

Solar or solar system abundance data derived from meteorites and the solar photosphere are reviewed periodically. The following is a (not necessarily complete) list of compilations that summarize information on photospheric and meteoritic abundances used as solar system abundance standards since 1989. For further reference in the following, a letter and number code is defined here for some of these compilations: Anders & Grevesse 1989 (A89), Palme & Beer 1993, Grevesse, Noels, & Sauval 1996, Grevesse & Sauval 1998 (GS98), Lodders 2003 (L03), Palme & Jones 2003 (PJ03), Asplund et al. 2005 (A05), Grevesse et al. 2007 (G07), and Lodders, Palme, & Gail 2009 (LPG09).

## 2 Meteorites as Abundance Standards for Non-Volatile Solar System Matter

About a century ago, it became more evident that meteorites may contain chemical and mineralogical information about the earliest solid objects that existed in the solar system, and that they may carry resemblance to the materials that accreted to planets like the Earth.

Several different meteorite groups are recognized. With respect to abundances, the chondritic meteorites are the most important ones. Meteorites that contain small silicate spheres are called chondrites (after Greek "chondros" for sphere). In addition to silicate minerals, most chondrites contain FeNi metal and iron sulfide (the mineral troilite, FeS, is most common), and a host of minor minerals. The major chondrite groups are ordinary, enstatite and carbonaceous chondrites; each group has further divisions based on the different proportions of their major minerals. The most common meteorites are, as the name implies, the "ordinary" chondrites. The enstatite chondrites, named after their most common silicate mineral enstatite, contain exotic sulfides and metal phases that are only stable under highly reducing conditions. The carbonaceous chondrites contain up to several percent of carbon, depending on sub-type. The carbonaceous meteorites of Ivuna type (abbreviated as CI chondrites for *C*arbonaceous and *I*vuna) are the most important ones for solar system abundance determinations.

The CI chondrites experienced severe aqueous alteration on their parent asteroid, and if they ever contained chondrules, metal, and sulfides as the other carbonaceous chondrites do, these phases were largely erased. The CI chondrites mainly consist of fine grained, hydrous silicates, magnetite ($Fe_3O_4$) probably produced by oxidation of FeNi metal, Fe-Ni bearing sulfides other than troilite, and various salts. Despite the absence of chondrules, the CI chondrites are still "chondrites" because their overall elemental composition for many elements is much closer to chondritic meteorites than to iron meteorites or so-called achondrites, that mainly consist of silicates and experienced severe melting.

Up to the 1950s, the CI chondrites were not given too much attention for their potential role as abundance standards. The classical abundance papers by e.g., Goldschmidt 1937, Brown 1949, and Suess & Urey 1956 used elemental analyses of silicates, metal, and sulfides from different chondrites and assumed some representative proportion of metal, silicate, and sulfides to come up with a meteoritic abundance standard.

By the 1970s, the picture emerged that the CI chondrites as well as another carbonaceous chondrite group named "CM" after the Mighei meteorite may be suitable groups to represent the solar system abundances of elements that provide the cations in rock-forming minerals. However, the relative contents of volatile elements is lower in CM chondrites than in CI chondrites, and a correlation of CM- to CI chondrite abundance ratios with condensation temperatures implies that volatility-related fractions occurred in CM chondrites.

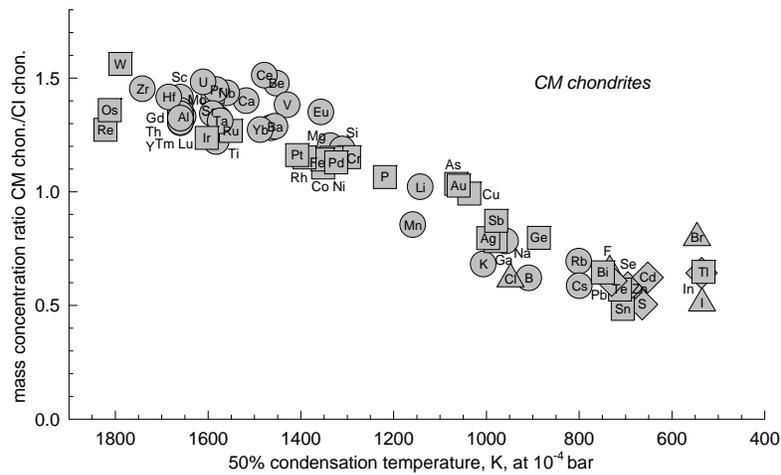

**Fig. 1.** The decreasing element concentration ratio of CM- over CI chondrites indicates volatility related fractionations in CM meteorites and makes them of limited use as an abundance standard. The different symbol shapes indicate the principal mineral host phase for the elements (circle: lithophile elements in silicate and oxides; box: siderophile elements in metal alloy; chalcophile: sulfides; triangle: halogen). Data sources for CM chondrites: Lodders & Fegley 1998 plus updates; CI chondrites: LPG09.

Figure 1 shows the concentration ratios for CM chondrites to CI chondrites as a function of condensation temperature. The symbols indicate the mineral phase hosting the elements. There is a smooth decrease in the concentration ratio with condensation temperatures which is independent of mineral host phase. This correlation indicates that the elemental abundances in CM chondrites are volatility controlled. The higher concentration ratio for refractory elements plots above unity and reflects that CM chondrites accumulated a higher proportion of refractory elements. The ratio below unity for the volatile elements incomplete condensation or accumulation. This limits the use of CM chondrites as abundance standards for elements with condensation temperatures less than ~ 1500 K. (see chapter by B. Fegley & L. Schaefer in this volume for condensation chemistry of the elements).

The "EH" enstatite chondrite group also has higher relative abundances of volatile elements, however, in detail chemical fractionations of volatile and non-volatile elements are apparent. In a comparison of the abundances of all chondrite groups to those in the solar photosphere, only the CI chondrites are found to have the closest match. Finally, the CI chondrites give the best agreement (of all chondrite groups) between observed and theoretically predicted nuclide abundances as a function of mass number (see e.g., Anders 1971 for arguments in favor of CI chondrites as standard).

Part of the reason why it took so long to recognize the significance of CI chondrites is simply that CI chondrites are very rare. Out of the ~1000 recorded observed meteorite falls from which material is preserved, only 5 CI chondrites are known. Among the 40,000 or so meteorites collected in Antarctica, only a few are CI chondrites. The meteorites are very fragile and decompose easily, for example, if placed into water, CI chondrites immediately begin to disintegrate. Hence, CI chondrites that are found a long time after their fall are not useful for abundance studies, as chemical information is easily altered or lost.

Table 1 lists the 5 observed CI chondrite falls, which are named after the nearest town to their fall location. Sufficient mass for study is only available for three of them, notably the Orgueil meteorite, which is probably one of the most and best analyzed rocks on this planet.

**Table 1**. Observed CI chondrite meteorite falls

| Meteorite | Date of Fall | Country | Preserved Mass |
|---|---|---|---|
| Alais | 15 March 1806 | France | 6 kg |
| Orgueil | 14 May 1868 | France | 14 kg |
| Tonk | 22 Jan. 1911 | India | 10 g |
| Ivuna | 16 Dec. 1938 | Tanzania | 0.7 kg |
| Revelstoke | 31 March 1965 | Canada | <1 g |

The other reason why it took so long to realize that CI chondrites are a chemically special meteorite group is that early chemical analysis methods were not able to analyze trace element abundances in small samples with sufficient precision. The advances came through the application of instrumental neutron activation analyses and mass-spectroscopic measurements, which revealed the chemical differences in minor element abundances. Compared to all other chondrites, the CI chondrites have the highest relative contents of volatile elements, such as C, O, alkalis, chalcogenides, and halogens. In contrast, the abundances of refractory elements such as Al, Ca, Si, Mg, Fe, alkaline earths, and rare earth elements are less variable among chondrites, but important abundance differences for these elements exist between the chondrite groups and provide lively discussions among meteoriticists.

At the same time that elemental abundance analyses improved for rock samples, spectroscopic abundance determinations for the solar photosphere advanced. A comparison of elemental abundances in the solar photosphere to CI chondrites then showed the best agreement for most elements; the exceptions being H, C, N, O, and the noble gases. These elements form extremely volatile compounds that may have never accreted to the CI chondrite parent asteroid or were easily lost from CI chondrite material while in space or in the terrestrial environment.

## 2.1 Composition of CI Chondrites

The reference composition of CI chondrites in Table 2 is based on a new review by Lodders, Palme & Gail (2009; henceforth LPG09). The previous evaluations of the composition of CI chondrites were done by Lodders (2003; henceforth L03) and Palme & Jones (2003; henceforth PJ03). These two evaluations used slightly different approaches to derive recommended CI chondrite compositions. PJ03 emphasized the use of the Orgueil meteorite as the CI standard rock, because it is the most massive of the 5 CI chondrite falls and therefore the most analyzed one (see Table 1). L03 used data from all CI chondrites and computed weighted average compositions. However, since most analytical data are for the Orgueil meteorite, these weighted averages are also dominated by this chondrite. A comparison of data from Orgueil and the other four CI chondrites in L03 shows that compositional differences among CI chondrites are relatively small.

In our new CI chondrite compilation (LPG09), we included many new data on trace elements that have become available in recent years, primarily because of improvements in instrumentation. In particular, application of Inductively Coupled Plasma Mass Spectrometry (ICP-MS) led to many new high quality analyses. These data and resulting updates of CI chondrite compositions are discussed in LPG09, however, an extensive discussion of the database containing 200+ references on CI chondrite analyses is deferred to the future (Lodders & Palme, 2009).

In addition to computing average concentrations from several reliable analyses for a given element, we employed the "element ratio method" to find standard values for element concentrations in CI chondrites. This method relies on the fact that concentration *ratios* of elements with similar cosmo- and geochemical properties are usually more constant than the absolute measured element concentration. The reason for using this method is that analytical data for CI meteorites can show variable absolute concentrations for some elements, which is caused by the variable water contents and/or massive alteration of fine grained matrix material. In addition, small scale heterogeneities naturally arise because these meteorites are breccias (see Greshake et al. 1998, Morlok et al. 2006). If there are compositional heterogeneities in the meteorite, the measured concentration of elements residing in common mineral phases may vary from sample to sample, however, the concentration ratios of these elements relative to each other will remain constant. Therefore, it is sometimes practical to take the concentration ratio of a pair of geochemically similar elements from several samples and to use the absolute abundance of the accurately determined element of that pair to calculate the concentration of the other element. Since there are only three CI chondrites for which larger numbers of analysis exist, it is also useful to compare the element concentration ratios in CI chondrites to that of other carbonaceous chondrites. The inclusion of data, e.g., from CM chondrites, which are relatively closely related to the CI chondrites, increases the statistics for abundance determinations by the ratio method. However, in using other data than for CI chondrites, the explicit assumption is made that chemical fractionations of the elements are absent between CI chondrites and the other chondrite groups invoked. This assumption is usually justified as long as CM chondrites and refractory elements are involved. Figure 1 shows that the concentration ratio of refractory elements in CI and CM chondrites plots at a comparable constant level.

The element ratio method and the direct average method usually give consistent results within 3% for most elements, which is easily within the estimated uncertainty from the statistics or individual analyses. However, including recent new analyses, the data spread has become wider for the elements Y, Zr, Hf, Nb and Ta. This spread seems to indicate real heterogeneities for these elements in CI chondrite samples, and the ratio method may lead to more reliable results (see LPG09 for more details).

The selected concentrations for the Orgueil meteorite, considered as most representative for the CI chondrites, are listed in Table 2; details about the data sources are in LPG09. Concentrations are given as parts per million (ppm) by mass (10,000 ppm = 1 mass%). Corresponding atomic abundances normalized to $10^6$ Si atoms (the "cosmochemical abundance scale") are listed as well.

**Table 2.** CI chondrite composition

| Z | | ppm | σ | Si=$10^6$ | σ |
|---|---|---|---|---|---|
| 1 | H | 19700 | 2000 | 5.13E+06 | 5.1E+05 |
| 2 | He | 0.00917 | | 0.601 | |
| 3 | Li | 1.47 | 0.19 | 55.6 | 7.2 |
| 4 | Be | 0.0210 | 0.0015 | 0.612 | 0.043 |
| 5 | B | 0.775 | 0.078 | 18.8 | 1.9 |
| 6 | C | 34800 | 3500 | 7.60E+05 | 7.6E+04 |

**Table 2.** CI chondrite composition

| Z  |    | ppm       | σ      | Si=10$^6$ | σ       |
|----|----|-----------|--------|-----------|---------|
| 7  | N  | 2950      | 440    | 55300     | 8300    |
| 8  | O  | 459000    | 46000  | 7.63E+06  | 7.6E+05 |
| 9  | F  | 58.2      | 8.7    | 804       | 121     |
| 10 | Ne | 1.80E-04  |        | 0.00235   |         |
| 11 | Na | 4990      | 250    | 5.70E+04  | 2.8E+03 |
| 12 | Mg | 95800     | 2900   | 1.03E+06  | 3E+04   |
| 13 | Al | 8500      | 260    | 8.27E+04  | 2.5E+03 |
| 14 | Si | 107000    | 3000   | 1.00E+06  | 3E+04   |
| 15 | P  | 967       | 97     | 8190      | 820     |
| 16 | S  | 53500     | 2700   | 4.38E+05  | 2.2E+04 |
| 17 | Cl | 698       | 105    | 5170      | 780     |
| 18 | Ar | 0.00133   |        | 0.00962   |         |
| 19 | K  | 544       | 27     | 3650      | 180     |
| 20 | Ca | 9220      | 461    | 60400     | 3000    |
| 21 | Sc | 5.9       | 0.3    | 34.4      | 1.7     |
| 22 | Ti | 451       | 36     | 2470      | 200     |
| 23 | V  | 54.3      | 2.7    | 280       | 14      |
| 24 | Cr | 2650      | 80     | 13400     | 400     |
| 25 | Mn | 1930      | 58     | 9220      | 280     |
| 26 | Fe | 185000    | 6000   | 8.70E+05  | 2.6E+04 |
| 27 | Co | 506       | 15     | 2250      | 70      |
| 28 | Ni | 10800     | 300    | 4.83E+04  | 1.4E+03 |
| 29 | Cu | 131       | 13     | 541       | 54      |
| 30 | Zn | 323       | 32     | 1300      | 130     |
| 31 | Ga | 9.71      | 0.49   | 36.6      | 1.8     |
| 32 | Ge | 32.6      | 3.26   | 118       | 12      |
| 33 | As | 1.74      | 0.16   | 6.10      | 0.55    |
| 34 | Se | 20.3      | 1.42   | 67.5      | 4.7     |
| 35 | Br | 3.26      | 0.49   | 10.7      | 1.6     |
| 36 | Kr | 5.22E-05  |        | 1.64E-04  |         |
| 37 | Rb | 2.31      | 0.16   | 7.10      | 0.50    |
| 38 | Sr | 7.81      | 0.55   | 23.4      | 1.6     |
| 39 | Y  | 1.53      | 0.2    | 4.52      | 0.45    |
| 40 | Zr | 3.62      | 0.4    | 10.4      | 1.0     |
| 41 | Nb | 0.279     | 0.028  | 0.788     | 0.079   |
| 42 | Mo | 0.973     | 0.097  | 2.66      | 0.27    |
| 44 | Ru | 0.686     | 0.041  | 1.78      | 0.11    |
| 45 | Rh | 0.139     | 0.014  | 0.355     | 0.035   |
| 46 | Pd | 0.558     | 0.028  | 1.38      | 0.07    |
| 47 | Ag | 0.201     | 0.010  | 0.489     | 0.024   |
| 48 | Cd | 0.674     | 0.047  | 1.57      | 0.11    |
| 49 | In | 0.0778    | 0.0054 | 0.178     | 0.012   |
| 50 | Sn | 1.63      | 0.24   | 3.60      | 0.54    |
| 51 | Sb | 0.145     | 0.021  | 0.313     | 0.047   |
| 52 | Te | 2.28      | 0.16   | 4.69      | 0.33    |
| 53 | I  | 0.53      | 0.11   | 1.10      | 0.22    |
| 54 | Xe | 1.74E-04  |        | 3.48E-04  |         |
| 55 | Cs | 0.188     | 0.009  | 0.371     | 0.019   |

**Table 2.** CI chondrite composition

| Z | | ppm | σ | Si=10$^6$ | σ |
|---|---|---|---|---|---|
| 56 | Ba | 2.41 | 0.14 | 4.61 | 0.28 |
| 57 | La | 0.242 | 0.012 | 0.457 | 0.023 |
| 58 | Ce | 0.622 | 0.031 | 1.17 | 0.06 |
| 59 | Pr | 0.0946 | 0.0066 | 0.176 | 0.012 |
| 60 | Nd | 0.471 | 0.024 | 0.857 | 0.043 |
| 62 | Sm | 0.152 | 0.008 | 0.265 | 0.013 |
| 63 | Eu | 0.0578 | 0.0029 | 0.100 | 0.005 |
| 64 | Gd | 0.205 | 0.010 | 0.342 | 0.017 |
| 65 | Tb | 0.0384 | 0.0027 | 0.063 | 0.004 |
| 66 | Dy | 0.255 | 0.013 | 0.412 | 0.021 |
| 67 | Ho | 0.0572 | 0.0040 | 0.091 | 0.006 |
| 68 | Er | 0.163 | 0.008 | 0.256 | 0.013 |
| 69 | Tm | 0.0261 | 0.0018 | 0.041 | 0.003 |
| 70 | Yb | 0.169 | 0.008 | 0.256 | 0.013 |
| 71 | Lu | 0.0253 | 0.0013 | 0.038 | 0.002 |
| 72 | Hf | 0.106 | 0.005 | 0.156 | 0.008 |
| 73 | Ta | 0.0145 | 0.0015 | 0.0210 | 0.0021 |
| 74 | W | 0.0960 | 0.0096 | 0.137 | 0.014 |
| 75 | Re | 0.0393 | 0.0039 | 0.0554 | 0.0055 |
| 76 | Os | 0.493 | 0.039 | 0.680 | 0.054 |
| 77 | Ir | 0.469 | 0.023 | 0.640 | 0.032 |
| 78 | Pt | 0.947 | 0.076 | 1.27 | 0.10 |
| 79 | Au | 0.146 | 0.015 | 0.195 | 0.019 |
| 80 | Hg | 0.350 | 0.070 | 0.458 | 0.092 |
| 81 | Tl | 0.142 | 0.011 | 0.182 | 0.015 |
| 82 | Pb | 2.63 | 0.18 | 3.33 | 0.23 |
| 83 | Bi | 0.110 | 0.010 | 0.138 | 0.012 |
| 90 | Th | 0.0310 | 0.0025 | 0.0351 | 0.0028 |
| 92 | U | 8.10E-03 | 6.5E-04 | 8.93E-03 | 7.1E-04 |

## 3 Photospheric Abundances

In 1929, Russell reported the earliest comprehensive analysis of the solar photosphere for 56 elements. Since then, numerous element abundance data have been derived by spectroscopy of the solar photosphere.

Converting the absorption lines into abundances requires knowledge of line positions of neutral and ionized atoms, as well as their transition probabilities and lifetimes of the excited atomic states. In addition, a model of the solar atmosphere is needed. In the past years, atomic properties have seen many experimental updates, especially for the rare earth elements (see below). Older solar atmospheric models used local thermodynamic equilibrium (LTE) to describe the population of the quantum states of neutral and ionized atoms and molecules according to the Boltzmann and Saha equations. However, the ionization and excitation temperatures describing the state of the gas in a photospheric layer may not be identical as required for LTE. Models that include the deviations from LTE (=non-LTE) are used more frequently, and deviations from LTE are modeled by including treatments for radiative and collision processes (see, e.g., Holweger, 2001, Steffen & Holweger 2002).

Solar atmospheric models have evolved from one dimension to more complicated 2D and 3D models designed to take into account effects of convection and granulation on radiative transfer in the solar atmosphere. Recent applications of 3D models instead of older 1D models leads to lower abundances of several elements, notably oxygen, from previously determined values. Significant reductions of photospheric abundances in other elements (e.g., Na, Al, Si) were also found (see, e.g., A05). However, different 3D model assumptions lead to different results, see for example the discussion by Shi et al. (2008) for silicon. Hence abundances derived with 3D models still have to be regarded with some caution until model assumptions and details are sorted out.

Lower abundances for some elements derived from these models also cause problems for standard solar models that describe the evolution of the sun to its current radius and luminosity (see Basu & Anita 2008). Another problem is that some of the 3D abundances compare worse to meteoritic data than before. In the following preference is given to elemental abundances derived with more conservative solar atmospheric models; however, for some elements (e.g., P, S, Eu) the results from 1D/2D and some 3D models produce consistent results.

The solar photospheric abundances from various literature sources are listed in Table 3. Several new measurements have become available since the compilation by L03 that lists references to data not described below.

Elemental abundances are normalized to $10^{12}$ atoms of hydrogen. The ratio of the number of atoms of an element N(X), relative to the number of hydrogen atoms, N(H) is given on a logarithmic scale, and frequently used notations are:

$$A(X) = \log N(X)/N(H) + 12 = \log \varepsilon X$$

Uncertainties for the logarithmic scale are given in logarithmic units "dex" equivalent to an uncertainty factor on a linear scale. The uncertainty relation is $\sigma(\text{in \%}) = (10^{\sigma \text{ (in dex)}} - 1) \times 100$.

As already mentioned, most elements are determined by absorption spectroscopy. Exceptions are rare gases and some elements that have no accessible or only heavily blended lines for use in quantitative spectroscopy of the photosphere (e.g., As, Se, Br, Te, I, Cs). Sunspot spectra are used for e.g., F, Cl, In, and Tl, however, the abundance uncertainty for these elements is rather large. For the noble gases, theoretical considerations or data from other objects must be used to obtain representative "solar" values (see below and L03). The following describes updates for several elements

**Helium**: The He discovery was made from spectral lines in the coronal spectrum during a solar eclipse in 1868; however, despite being first discovered in the sun, the He abundance cannot be determined spectroscopically in the solar photosphere. The He abundance is determined from results of helioseismic models, as described below in the section for the present-day mass fractions of H, He, and heavy elements.

**Lithium**: The value of A(Li) = 1.1±0.1 from Carlsson et al. 1994 used in L03 is kept. The analysis from Mueller et al. 1975 of A(Li)=1.0±0.1 was amended with 3D models by A05 to give 1.05±0.10. Considering the already large uncertainty and the uncertainties in the 3D models, the previous selection is kept.

**Beryllium**: The photospheric and meteoritic Be abundance determinations are associated with difficulties (L03). By 2003, there were two conflicting photospheric determinations (A(Be)= 1.15±0.10 and 1.40±0.09). A subsequent re-analysis of the photospheric Be abundance by Asplund (2004) yields A(Be) = 1.38±0.09, in support of the higher photospheric value. The meteoritic value of A(Be) = 1.41±0.08 in L03 was based on chemical element systematics in CM and CV chondrites because there was only one previous determination of Be in CI chondrites. This situation has improved since Makishima & Nakamura (2006) report 2 Be analyses for the Orgueil CI chondrite. The average of the 3 Be determinations is 0.021±0.002 ppm (by mass), corresponding to a meteoritic A(Be)= 1.32±0.03.

Within the larger uncertainties, the photospheric and meteoritic Be abundances are in agreement, suggesting that no Be destruction has occurred over the Sun's lifetime. However, taken at face value, a lower Be abundance in CI chondrites than in the photosphere is not easily accounted for by any physical or chemical fractionation process.

**Carbon**: The C abundance of A(C) = 8.39(±0.04) from Allende Prieto et al. 2002 was selected in L03. This value was confirmed by Asplund et al. 2005b, and derived from the analysis of CO by Scott et al. 2006. However, this C abundance is based on 3D models from one group and an independent confirmation of this value by another model is desirable.

**Nitrogen**: The N abundance is among the more uncertain elemental abundances. A05 derived N abundances from neutral N (NI) and the NH molecule. An LTE analysis of the NI lines gives A(N) = 7.88±0.08, a NLTE analysis gives 7.85±0.08. The N abundance from the NH molecule and 3D models gives a lower abundance of A(N) = 7.73±0.05. A05 recommend A(N) = 7.78±0.06 in their table. However, this seems quite low in comparison to previous estimates of e.g., 7.83±0.11 in L03 that was based on Holweger 2001. A recent detailed study by Caffau et al. (2009) found 7.86±0.12, which is recommended here.

**Oxygen**: The recommended O abundance of 8.73±0.07 is an average from the recent determinations by Caffau et al. 2008a, Ludwig & Steffen 2007, and Melendez & Asplund 2008.

Caffau et al. 2008a recommend A(O)= 8.76 ±0.07 from a detailed analyses of several oxygen lines and 3D atmospheric models. They include different model NLTE corrections, as well as treatments of collisions with H atoms on NLTE level populations of O. Within the (unfortunately still larger) uncertainties, their O abundance is closer to most of the recent low O abundance determinations than to the older values advocated by AG989 or GS98. Melendez & Asplund 2008 derived the O abundances from several O lines and compared various model approaches, which lead to an average O abundance of A(O) = 8.71±0.02, where the uncertainty does not really cover the real uncertainty of the "true" O abundance. Using different 3D models, Ludwig & Steffen 2007 report A(O) = 8.72±0.06 from modeling several O I lines.

A review on the problem of the solar O and other light element abundances is given by Basu & Anita 2008. The photospheric O abundance determination remains enigmatic, as the value obtained by Allende-Pietro et al. 2001 and Asplund et al. 2004 appears to be erratically low, whereas the older abundances quoted in compilations such as AG89, GS98, as well as the recent study by Ayres et al. 2006 probably give values that are much too high.

The downward adjustment in solar O by Allende-Pietro et al. 2001 was substantial compared to the values given in the compilations by GS98 (A(O)= 8.83) and AG89 (A(O) = 8.93). The decrease was mainly due to the realization that the O line commonly used in the abundance analysis is blended with a Ni line. The value by Caffau et al. is 0.07 dex higher (factor 1.17) than the abundances derived by Allende-Pietro et al. (2001); and it is also higher than the low values determined by Asplund and coworkers. Caffau et al. (2008a) give these reasons why the O abundances was previously underestimated. In the studies by Allende Pietro et al. and Asplund et al., lower equivalent widths were used, and effects of collisions of H atoms in the calculations of the statistical equilibrium of O were not considered.

Ayres et al. (2006) derived the O abundance from weak CO absorptions but did not find support for a lower O abundance. They recommend A(O) = 8.85, much closer to the value in AG89 (A(O) = 8.93). However, to obtain the absolute O abundance using the CO molecule, Ayres et al. assumed that C/O = 0.5. Oxygen is about twice as abundant as carbon, and carbon is largely tied into the CO molecule. An analysis of the CO abundance thus provides only the lower limit to the total C abundance (as C is present also in other gases such as C, CH etc), and the O abundance can only be derived if the total C/O ratio and the C abundance is known. Ayres et al. 2006 find A(C) =8.54, and with an assumed C/O of 0.5, their corresponding O abundance is A(O) = 8.85. However, using the same CO lines Scott et al. 2006 find a lower O abundance with their 3D models than reported by Ayres et al. 2006. Given the problem that the O abundances from CO requires various assumptions about the distribution of C and the C/O ratio, these abundance determinations appear even more uncertain.

Adopting A(C) = 8.39 and A(O) = 8.73±0.07 gives a C/O ratio of C/O = 0.457, less than the C/O ratio of 0.50 in more recent studies and compilations, but still somewhat higher than the value of C/O = 0.427 from AG89.

The O abundance has been steadily revised downward from Russell's 1929 value over the years, likewise, the abundance values of C and N from more recent analysis tend to be smaller. Figure 2 shows historical abundance trends for the more abundant elements C, N, O, Si, and Fe from various sources starting with Russell 1929 and including data sources quoted in the compilations mentioned above. The variations on the graph appear small, however, one should not forget that the data are plotted on a log scale and a difference of 0.5 dex corresponds to about a factor of 3 change in abundance. The Fe abundance was a big problem until the early 1970s when improved transition probabilities and lifetimes confirmed that the photospheric Fe abundance had been underestimated by about a factor of 3-10, which did not match the meteoritic value. This issue has now been put to rest and the photospheric and meteoritic value are now in perfect agreement. However, the issue of the C, N, and O abundances cannot be aided with meteoritic abundances, and future analyses have to resolve this.

**Neon**: Results by Morel & Butler (2008) from Ne I and Ne II lines in nearby, early type B stars yields log A(Ne) = 7.97 ± 0.07, which can be taken as characteristic of the present day ISM. If neon contributions from more massive AGB stars to the ISM over the past 4.6 Ga are negligible, this value may be taken as representative for the Sun. Landi et al. 2007 obtained A(Ne) = 8.11±0.1 from solar flare measurements in the ultra-violet. This Ne abundance is derived independently of the O abundance, unlike other Ne abundance determinations that rely on Ne/O ratios and an adopted O abundance (see e.g., L03 and below). The average from these studies gives A(Ne) = 8.05±0.06, but considering the larger uncertainties in the determinations, an overall uncertainty of 0.10 dex (25%) is easily warranted for the recommended value here.

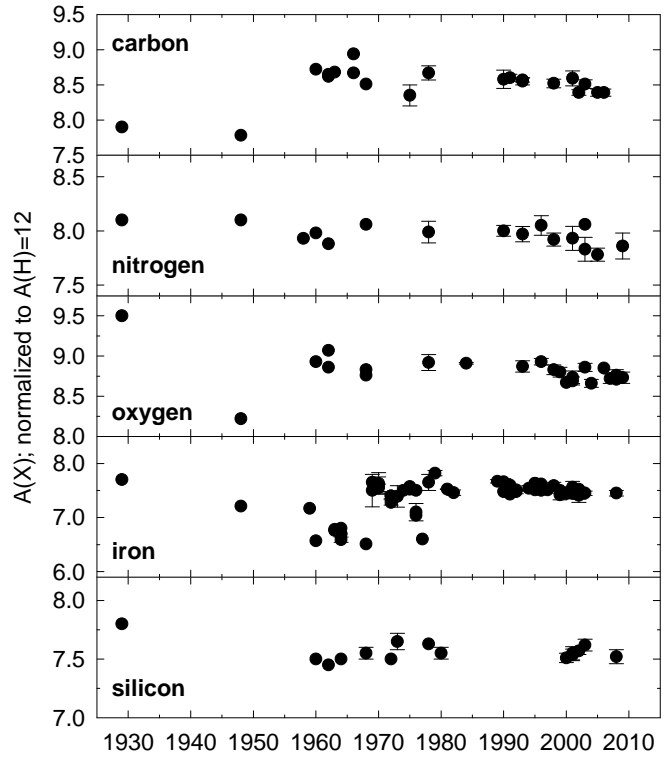

**Fig. 2.** Photospheric abundance determinations over time

This Ne value compares relatively well to the Ne abundance adopted in older compilations by AG89 (8.09) or GS98 (8.08). However, the Ne values in more recent compilations (A(Ne) = 7.87±0.1 in L03, and A(Ne) = 7.84±0.06 in A05) are lower by 26% (0.10 dex). The Ne abundance in previous compilations was mainly based on the characteristic Ne/O ratio of 0.15 for the local ISM and solar energetic particles. The Ne abundance was then calculated from the adopted O abundances. Since the adopted O abundances have changed to lower values in these compilations, the Ne abundances dropped as well. Using Ne/O = 0.15 and our selected O abundance from above, the Ne abundance from the ratio method gives A(Ne) = 7.91, about 0.14 dex lower than the preferred Ne value which is independent of an adopted O abundance. More recently, Young (2005) recommends Ne/O = 0.17±0.05 for the photosphere from extreme UV measurements of supergranule cell center regions. This ratio and the selected O abundance (A(O)=8.73) yields A(Ne) = 7.96 (±0.15, uncertainty only from ratio). This value is lower than the recommended value, but agrees within uncertainties permitted by the given Ne/O ratio.

In principle, the Ne abundance can also be derived from solar wind data. In 2006, Bochsler et al. reported A(Ne) 8.08±0.12. Bochsler et al. 2007 further analyzed the solar wind data for fractionations, and finds A(Ne) =7.96±0.13, and also A(O) = 8.87±0.11, with relatively large uncertainties that do not help to resolve the issue of the uncertain O and Ne abundances.

Neon is the third most abundant heavy element after O and C, and the heavy elements are important opacity sources that influence radiative transfer in the sun. Good agreement of standard solar models and helioseismological observations existed until about the year 2000. Then more sophisticated photospheric modeling began to yield lower C, N, O, Ne and other heavier element abundances (see Figure 2, and compilations by L03,A05). A decrease in heavy element abundances led to solar model results that no longer stood the test from helioseismology. A detailed review of this problem is given by Basu & Anita (2008). The recommended N, O, and Ne abundances here are larger than previously recommended in L03 and A05, and it will be interesting to see if these abundances can bring solar models again in closer agreement with helioseismological constraints.

**Sodium:** the previously selected Na value of 6.3±0.03 in L03 is in agreement with A(Na) = 6.27 measured by Reddy et al. (2003; henceforth R03). A05 found A(Na) = 6.17±0.04 from six Na lines and their 3D model atmospheres, which is ~25% lower than the meteoritic value as well as previously determined photospheric Na abundances. The reason for this difference is not yet clear.

**Aluminum:** A 3D analysis by A05 gives A(Al) = 6.37±0.06, which is lower than the previous value of 6.47±0.07 from 1D models selected in L03. The older value is kept as it is much better in agreement with meteoritic data. The Al/Si from 3D models in A05 is 1.2 times that of CI chondrites, which is a large difference that still needs to be understood.

**Silicon**: The A(Si)=7.52±0.06 selected here is from Shi et al. (2008), which is not that different from 7.51±0.04 reported by Asplund 2000 and 7.54±0.05 by Holweger 2001.

**Phosphorus**: The value of A(P)=5.49±0.04 in L03 is changed to the recent result of A(P) = 5.46±0.04 by Caffau et al. 2007a. The new value is based on 3D atmospheric models. According to Caffau et al. 2007a the value for P with 1D models is not significantly different. A lower value, A(P) = 5.36±0.04 from a different 3D analysis was given in the compilation by A05. Here the well documented analysis by Caffau et al. 2007a is taken for the photospheric abundance, which agrees well with the meteoritic value of 5.43±0.04.

**Sulfur**: New results with 3D models by Caffau & Ludwig 2007 and Caffau et al. 2007b lead to A(S) = 7.14±0.01. They found that 3D models have no big effect on S abundances compared to 1D models.

**Argon**: The value of A(Ar) = 6.50 is derived from various independent sources since the Ar abundance cannot be determined spectroscopically in the photosphere (see Lodders 2008).

**Potassium**: Zhang et al. 2006 confirmed the K abundance of A(K) = 5.12±0.03 used in L03. The value A(K)=5.08±0.07 proposed by A05 appears too low, a similar situation as for Na above.

**Calcium**: Reddy et al. 2003 find A(Ca) = 6.33±0.07 in their LTE analysis of the solar spectrum (the same value as Lambert's 1968 one), which is lower than the previously selected value in L03. However, the uncertainty for the value from R03 is high. The result from 3D models by A05 is about 5% lower. A new Ca analysis for the photosphere using different model atmospheres is needed.

**Scandium**: The Sc abundance remains uncertain. Zhang et al. (2008) recommend a range of 3.07 < A(Sc) < 3.13. To cover the range of values reported in several recent papers A(Sc) = 3.10 with an appropriate uncertainty of 0.1 dex is recommended.

**Titanium**: The value from the LTE analysis by R03 is adopted.

**Chromium**: Sobeck et al. (2007) found A(Cr) = 5.64±0.1, which is identical with the photospheric value listed in L03 but with much smaller uncertainty.

**Manganese**: Two recent studies give A(Mn) = 5.37±0.05 (Bergemann & Gehren 2007) and 5.36±0.10 (Blackwell-Whithhead & Bergemann, 2007), and R03 found the same value as Bergemann & Gehren (2007). These values are only slightly lower than the value 5.39±0.03 given in L03. The analyses of the photospheric Mn abundance seem to confirm that the photospheric Mn abundance is lower than the meteoritic value of 5.50±0.01. Assuming that indeed both the photospheric and meteoritic data are reliable, the cause of this abundance difference must be found.

**Nickel**: The A(Ni) = 6.23±0.04 from the LTE analysis by R03 is similar to the previous value and has a lower uncertainty.

**Zirconium**: Ljung et al. 2006 found A(Zr) = 2.58±0.02 from a 3D analysis.

**Palladium**: A value of A(Pd) = 1.66±0.04 was reported by Xu et al. 2006

**Indium**: Previous determinations of the photospheric In abundance lead to a value that is substantially higher (~400%) than the CI chondritic value. Gonzales (2006) suggested that this large difference could be the result of the relatively high volatility of In. Incomplete condensation of any element into the materials assembled to the CI chondrite parent body would lead to a relative depletion of a volatile element in CI chondrites when compared to the Sun. At $10^{-4}$ bar total pressure, half of all In is condensed at 536 K, comparable to the 50% condensation temperatures of other volatile elements like Tl (532K), S (664K) and Pb (730K; see L03). However, the abundances of these and other volatile elements are in closer agreement for the photosphere and CI chondrites, so a fractionation due to volatility cannot explain the huge photospheric In abundance.
 A high In abundance can also be ruled out considering the abundance distribution of the elements and nuclides as a function of atomic mass. A high In abundance would introduce "spikes" in the otherwise rather smooth abundances curve (see below). A recent investigation by Vitas et al. (2008) shows that the often used In line at 451.13 nm in the solar sun spot spectrum is blended by some line of a currently not identified element, which causes an apparently higher In abundance. Vitas et al. (2008) obtain A(In) = 1.50 if no potential blends are considered, and this value is adopted here as an upper limit for the photospheric In abundance. Their study of other In lines further indicates that the solar In abundance is unlikely to be higher than the meteoritic value of A(In) = 0.78, for which Vitas et al. (2008) also find support from nuclide distribution systematics from nucleosynthesis.

**Rare earth elements**: In the past years, many improvements have been made in abundance analyses of the REE through measurements of atomic lifetimes and transition probabilities, notably by the Wisconsin group. A recent paper by Sneden et al. 2009 summarizes the efforts and gives abundances for the REE. Now the REE abundances are among the best-known abundances for the sun. The following lists several papers on REE that appeared since 2003; the new values for Ce, Dy, Tm, Yb, and Lu are from Sneden et al. 2009.

**Pr**: The value of A(Pr)= 0.76±0.02 from Sneden et al. (2009) is preferred to the previously selected value of A(Pr)=0.71±0.08 in L03. The value based on Ivarsson et al. 2003 of A(Pr) = 0.58±0.10 adopted by G07 is erroneously low, which is easily seen from a comparison to the well-established meteoritic value of 0.78±0.03.

**Nd**: Den Hartog et al. 2003; A(Nd)=1.45±0.05

**Sm**: Lawler et al. 2006b; A(Sm)=1.00±0.03

**Eu**: Mucciarelli et al. (2008) employ a 3D hydrodynamic model atmosphere and find A(Eu) = 0.52 ± 0.02. Mucciarelli et al. find that 3D effects are negligible for the Eu determination in the Sun. This value is identical to the value reported by Lawler et al. 2001.

**Gd**: Den Hartog et al. 2006; A(Gd) = 1.11±0.03.

**Ho**: Lawler et al. 2004; A(Ho) = 0.51±0.1.

**Er**: Lawler et al. 2008c; A(Er) = 0.96±0.03

**Hafnium**: The photospheric abundance of A(Hf) = 0.88±0.08 selected in L03 is confirmed with a reanalysis by Lawler et al. 2007 who determined improved transition probabilities. Another recent re-determination of the photospheric Hf abundance gives A(Hf) = 0.87±0.04 (Caffau et al. 2008b).

**Osmium**: Quinet et al. 2006 found A(Os) = 1.25±0.11. This is significantly lower than the value of 1.45±0.10 used in previous compilations. Both Os values seem to be problematic when compared to the meteoritic value of 1.37±0.03. Assuming that the meteoritic value is reliable, the older photospheric value is 17% too low and the new one 30% too high. Grevesse et al. 2007 selected the newer, smaller value but a conservative approach is to adopt the value with the smaller difference to the meteoritic value. The older value of A(Os) = 1.45 also appears more reasonable considering abundance systematics in the Pt-element region. Nucleosynthesis models predict that Os should be more abundant than Ir, as seen in CI chondrites. Overall, a new analysis of the photospheric Os abundance is needed.

**Platinum**: Den Hartog et al. 2005 find that the photospheric Pt abundance is not very reliable. The value selected in L03 is kept here, but is assigned a 0.3 dex uncertainty (factor of 2) to emphasize its low reliability.

**Thallium**: The Tl value of 0.95±0.2 is from the linear average of the endmember composition of $0.72 \leq A(Tl) \leq 1.1$ selected in L03 that was found for sunspot spectra. The uncertainty quoted here is to indicate the derived range. There are no new measurements.

**Thorium**: The Th abundance is difficult to determine because the only accessible Th line in the photospheric spectrum is heavily blended with Ni I and Ni II. Caffau et al. 2008b report a nominal Th abundance $A(Th) = 0.08\pm0.03$, which should not be over-interpreted because of the line blends.

## 4 Recommended Present-Day Solar Abundances

### 4.1 Cosmochemical and Astronomical Abundance Scale Conversion

In order to compare the atomic silicon-normalized CI chondrite abundances in Table 2 ($N(Si) = 10^6$ atoms; cosmochemical abundance scale) with the photospheric abundances on the hydrogen-normalized scale ($A(H) = 12$; astronomical abundance scale) in Table 3, the data must be converted to a common scale. One cannot easily convert the meteoritic data to the H-normalized astronomical abundance scale because H is depleted in meteorites. However, a comparison can be done for the non-volatile rock-forming elements. The difference of the logarithmic Si-normalized abundances of CI chondrites to the abundances on the astronomical scale is more or less constant for many elements. This shows that the relative abundances in the photosphere and CI chondrites are similar.

The link for both abundance scales in an average conversion constant that is calculated by subtracting the logarithm of the Si-normalized meteoritic abundances (Table 2) from the logarithmic H-normalized photospheric abundances (Table 3) for all elements heavier than neon that have uncertainties < 0.1 dex, i.e., below ~25%, in their photospheric abundance determinations. There are 39 elements that qualify and Figure 3 shows a comparison of the photospheric and CI chondritic abundances for these elements on a linear scale (note that the conversion constant for the log scales is equivalent to a scale factor on linear abundance scales). The scale conversion constant is 1.533 ± 0.042; and the cosmochemical and astronomical scales are coupled as:

$$A(X) = \log N(X) + 1.533$$

Previously, the conversion constant was somewhat larger. For example, AG89 used 1.554 ± 0.020 from only 12 elements, which resulted in a smaller nominal uncertainty of the conversion constant. Lodders 2003 found a value of 1.539 ± 0.046 for 35 elements for which photospheric abundances were determined with less than 25% uncertainty, and used a constant of 1.540, which is exactly the log of the ratio of Si in the astronomic to the meteoritic scale. The slightly lower conversion factor found here is the result of the systematic decrease of the reported photospheric abundance values.

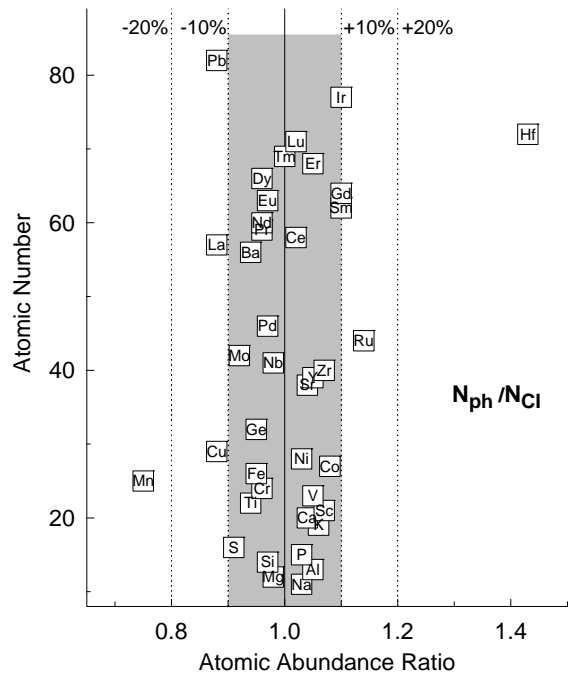

**Fig. 3.** The photospheric/CI chondritic abundance ratios for 39 elements that are well determined for the solar photosphere. The grey-shaded region shows agreement within 10%.

It is notable that the conversion constant can be derived from a large range of elements with different properties (e.g., atomic number, mass, first ionization potential, condensation temperature). The premise in linking the meteoritic and solar data is that there are no chemical and physical fractionations of the elements (except for the obvious loss of highly volatile elements from meteorites). The small spread in the conversion factor indicates that there is basic agreement of solar and meteoritic abundances. There is no apparent dependence of the conversion factor on atomic number, mass or any other elemental property. In addition, the solar/meteoritic abundance ratios are independent of the geochemical character of an element, whether it is lithophile, siderophile or chalcophile, which indicates that any chemical and physical fractionation of silicates, metal, and sulfides did not affect CI chondrite abundances. A reasonable estimate for the uncertainty of the relative scale of solar and meteoritic abundances is about 10%.

**4.2 Comparison of Photospheric and Meteoritic Abundances**

The photospheric and CI chondrite abundances on the astronomical abundance scale are given in Table 3. Agreement within 10% for meteoritic and photospheric data exists for 40 elements (the 39 shown in Figure 3 plus the light element Be). This increase in elements that show good agreement is mainly due to the recent improvements in photospheric measurements.

The largest differences are for the highly volatile elements that form low-temperature ices and/or exist in gaseous form in the terrestrial atmosphere. The largest depletion is for the noble gases. The depletion sequence for N, C, H reflects the general lack of solid nitrogen compounds in meteorites and the predominance of oxides and silicates.

Only Li is clearly consumed in the interior of Sun by nuclear reactions, which explains the ~150 times smaller photospheric abundance. Nominally, B is depleted in the Sun by about 20 % but within the stated uncertainties it is apparently not affected. Beryllium is another fragile element like Li and B, and may be subject to destruction in the Sun. However, the comparison of abundances for the photosphere and CI chondrites indicate that there was no Be loss in the Sun; indeed; the nominal Be abundance for the Sun is higher than in CI chondrites.

The difference between photospheric and CI chondrite abundances exceeds 10% for 21 other elements (see LPG09 for a detailed comparison). However, in most cases the combined uncertainties of the photospheric and meteoritic determinations are larger than the difference in abundance, and solar and meteoritic abundances agree within error limits.

Elements with abundance differences larger than the combined error bars are W, Rb, Ga, Hf, and Mn. The abundances of Ga, Rb, and W need to be re-determined in the photosphere to resolve the differences. There are new photospheric analyses for Hf and Mn suggesting that the differences in photospheric and meteoritic values could be real since there no plausible reasons to doubt the results. Line blending may not be the culprit as this usually leads to over-estimated abundances for the photosphere (e.g., like for Indium as noted above); however, the photospheric Mn value is ~1.4x lower than in CI chondrites. Manganese can be accurately measured in meteorites. Its concentration in Orgueil is similar to that in two other CI meteorites, Alais and Ivuna (L03), and it fits with the abundances of other elements of similar volatility. It seems that an unidentified problem in the photospheric abundance analysis may cause the discrepancy of the meteoritic and solar Mn abundances.

The problem is reversed for Hf and the meteoritic abundance of Hf is less than that of the photosphere. This could indicate a problem with the photospheric abundance determination and suggest that line blending is more severe than already corrected for in current models. However, two recent Hf analyses using different models essentially obtain the same abundance, and if there is a problem with the analysis, it remains elusive. The Hf concentration in CI chondrites has been accurately determined, because Hf is important for Lu-Hf and W-Hf dating. The very constant Lu/Hf ratio in meteorites closely ties Hf to other refractory elements, which do not show large differences in abundance to the sun as does Hf. This issue awaits resolution.

Overall, the agreement between photospheric and meteoritic abundances has improved further with new photospheric and meteoritic data.

## 4.3 Combined Solar Abundances from CI Chondrites and Photospheric Data

The CI chondritic and photospheric abundances can be combined to select a set of recommended present-day solar system abundances. Here the same procedure as in L03 is used to construct such an abundance set. The recommended data are from photospheric values for ultra-highly volatile elements like H, C, N, and O and from various sources and theory for the noble gases (see above for He, Ne, Ar, and L03 for Kr and Xe.). The CI chondrite data are the obvious choice for elements that are only determined in CI chondrites but also for elements that have photospheric abundance determinations with high uncertainties. Several elements are equally well determined in CI chondrites and in the photosphere, and an average of their Si-normalized abundances is used. The recommended present day abundances are converted to the astronomical scale using the same conversion constant (1.533) between the astronomical and cosmochemical abundance scales as described before.

**Table 3.** CI chondrite, solar (mainly photosphere), and recommended present-day solar abundances

| Z | | CI Chondrites A(X) | | Sun A(X) | | Note | Recommended A(X) | |
|---|---|---|---|---|---|---|---|---|
| 1 | H | 8.24 | 0.04 | 12.00 | | s | 12.00 | |
| 2 | He | 1.31 | | 10.925 | 0.02 | s,t | 10.925 | 0.02 |
| 3 | Li | 3.28 | 0.05 | 1.10 | 0.10 | m | 3.28 | 0.05 |
| 4 | Be | 1.32 | 0.03 | 1.38 | 0.09 | m | 1.32 | 0.03 |
| 5 | B | 2.81 | 0.04 | 2.70 | 0.17 | m | 2.81 | 0.04 |
| 6 | C | 7.41 | 0.04 | 8.39 | 0.04 | s | 8.39 | 0.04 |
| 7 | N | 6.28 | 0.06 | 7.86 | 0.12 | s | 7.86 | 0.12 |
| 8 | O | 8.42 | 0.04 | 8.73 | 0.07 | s | 8.73 | 0.07 |
| 9 | F | 4.44 | 0.06 | 4.56 | 0.30 | m | 4.44 | 0.06 |
| 10 | Ne | -1.10 | | 8.05 | 0.10 | s,t | 8.05 | 0.10 |
| 11 | Na | 6.29 | 0.02 | 6.30 | 0.03 | a | 6.29 | 0.04 |
| 12 | Mg | 7.55 | 0.01 | 7.54 | 0.06 | a | 7.54 | 0.06 |
| 13 | Al | 6.45 | 0.01 | 6.47 | 0.07 | a | 6.46 | 0.07 |
| 14 | Si | 7.53 | 0.01 | 7.52 | 0.06 | m | 7.53 | 0.06 |
| 15 | P | 5.45 | 0.04 | 5.46 | 0.04 | a | 5.45 | 0.05 |
| 16 | S | 7.17 | 0.02 | 7.14 | 0.01 | a | 7.16 | 0.02 |
| 17 | Cl | 5.25 | 0.06 | 5.50 | 0.30 | m | 5.25 | 0.06 |
| 18 | Ar | -0.48 | | 6.50 | 0.10 | s,t | 6.50 | 0.10 |
| 19 | K | 5.10 | 0.02 | 5.12 | 0.03 | a | 5.11 | 0.04 |
| 20 | Ca | 6.31 | 0.02 | 6.33 | 0.07 | m | 6.31 | 0.02 |
| 21 | Sc | 3.07 | 0.02 | 3.10 | 0.10 | m | 3.07 | 0.02 |
| 22 | Ti | 4.93 | 0.03 | 4.90 | 0.06 | m | 4.93 | 0.03 |
| 23 | V | 3.98 | 0.02 | 4.00 | 0.02 | a | 3.99 | 0.03 |
| 24 | Cr | 5.66 | 0.01 | 5.64 | 0.01 | a | 5.65 | 0.02 |
| 25 | Mn | 5.50 | 0.01 | 5.37 | 0.05 | m | 5.50 | 0.01 |
| 26 | Fe | 7.47 | 0.01 | 7.45 | 0.08 | a | 7.46 | 0.08 |
| 27 | Co | 4.89 | 0.01 | 4.92 | 0.08 | a | 4.90 | 0.08 |
| 28 | Ni | 6.22 | 0.01 | 6.23 | 0.04 | a | 6.22 | 0.04 |
| 29 | Cu | 4.27 | 0.04 | 4.21 | 0.04 | m | 4.27 | 0.04 |
| 30 | Zn | 4.65 | 0.04 | 4.62 | 0.15 | m | 4.65 | 0.04 |
| 31 | Ga | 3.10 | 0.02 | 2.88 | 0.10 | m | 3.10 | 0.02 |
| 32 | Ge | 3.60 | 0.04 | 3.58 | 0.05 | a | 3.59 | 0.06 |
| 33 | As | 2.32 | 0.04 | 0.00 | 0.00 | m | 2.32 | 0.04 |
| 34 | Se | 3.36 | 0.03 | 0.00 | 0.00 | m | 3.36 | 0.03 |
| 35 | Br | 2.56 | 0.06 | 0.00 | 0.00 | m | 2.56 | 0.06 |
| 36 | Kr | -2.25 | | 3.28 | 0.08 | t | 3.28 | 0.08 |
| 37 | Rb | 2.38 | 0.03 | 2.60 | 0.10 | m | 2.38 | 0.03 |

**Table 3.** CI chondrite, solar (mainly photosphere), and recommended present-day solar abundances

| Z | | CI Chondrites A(X) | | Sun A(X) | | Note | Recommended A(X) | |
|---|---|---|---|---|---|---|---|---|
| 38 | Sr | 2.90 | 0.03 | 2.92 | 0.05 | m | 2.90 | 0.03 |
| 39 | Y | 2.19 | 0.04 | 2.21 | 0.02 | a | 2.20 | 0.04 |
| 40 | Zr | 2.55 | 0.04 | 2.58 | 0.02 | a | 2.57 | 0.04 |
| 41 | Nb | 1.43 | 0.04 | 1.42 | 0.06 | a | 1.42 | 0.07 |
| 42 | Mo | 1.96 | 0.04 | 1.92 | 0.05 | a | 1.94 | 0.06 |
| 44 | Ru | 1.78 | 0.03 | 1.84 | 0.07 | m | 1.78 | 0.03 |
| 45 | Rh | 1.08 | 0.04 | 1.12 | 0.12 | a | 1.10 | 0.13 |
| 46 | Pd | 1.67 | 0.02 | 1.66 | 0.04 | a | 1.67 | 0.04 |
| 47 | Ag | 1.22 | 0.02 | 0.94 | 0.30 | m | 1.22 | 0.02 |
| 48 | Cd | 1.73 | 0.03 | 1.77 | 0.11 | m | 1.73 | 0.03 |
| 49 | In | 0.78 | 0.03 | 1.50 | UL | m | 0.78 | 0.03 |
| 50 | Sn | 2.09 | 0.06 | 2.00 | 0.30 | m | 2.09 | 0.06 |
| 51 | Sb | 1.03 | 0.06 | 1.00 | 0.30 | m | 1.03 | 0.06 |
| 52 | Te | 2.20 | 0.03 | 0.00 | 0.00 | m | 2.20 | 0.03 |
| 53 | I | 1.57 | 0.08 | 0.00 | 0.00 | m | 1.57 | 0.08 |
| 54 | Xe | -1.93 | | 2.27 | 0.08 | t | 2.27 | 0.08 |
| 55 | Cs | 1.10 | 0.02 | 0.00 | 0.00 | m | 1.10 | 0.02 |
| 56 | Ba | 2.20 | 0.03 | 2.17 | 0.07 | a | 2.18 | 0.07 |
| 57 | La | 1.19 | 0.02 | 1.14 | 0.03 | m | 1.19 | 0.02 |
| 58 | Ce | 1.60 | 0.02 | 1.61 | 0.06 | a | 1.60 | 0.06 |
| 59 | Pr | 0.78 | 0.03 | 0.76 | 0.04 | a | 0.77 | 0.05 |
| 60 | Nd | 1.47 | 0.02 | 1.45 | 0.05 | m | 1.47 | 0.02 |
| 62 | Sm | 0.96 | 0.02 | 1.00 | 0.05 | m | 0.96 | 0.02 |
| 63 | Eu | 0.53 | 0.02 | 0.52 | 0.04 | a | 0.53 | 0.04 |
| 64 | Gd | 1.07 | 0.02 | 1.11 | 0.05 | a | 1.09 | 0.06 |
| 65 | Tb | 0.34 | 0.03 | 0.28 | 0.10 | m | 0.34 | 0.03 |
| 66 | Dy | 1.15 | 0.02 | 1.13 | 0.06 | a | 1.14 | 0.06 |
| 67 | Ho | 0.49 | 0.03 | 0.51 | 0.10 | m | 0.49 | 0.03 |
| 68 | Er | 0.94 | 0.02 | 0.96 | 0.06 | a | 0.95 | 0.06 |
| 69 | Tm | 0.14 | 0.03 | 0.14 | 0.04 | m | 0.14 | 0.03 |
| 70 | Yb | 0.94 | 0.02 | 0.86 | 0.10 | m | 0.94 | 0.02 |
| 71 | Lu | 0.11 | 0.02 | 0.12 | 0.08 | m | 0.11 | 0.02 |
| 72 | Hf | 0.73 | 0.02 | 0.88 | 0.08 | m | 0.73 | 0.02 |
| 73 | Ta | -0.14 | 0.04 | 0.00 | 0.00 | m | -0.14 | 0.04 |
| 74 | W | 0.67 | 0.04 | 1.11 | 0.15 | m | 0.67 | 0.04 |
| 75 | Re | 0.28 | 0.04 | | | m | 0.28 | 0.04 |
| 76 | Os | 1.37 | 0.03 | 1.45 | 0.11 | m | 1.37 | 0.03 |
| 77 | Ir | 1.34 | 0.02 | 1.38 | 0.05 | a | 1.36 | 0.06 |
| 78 | Pt | 1.64 | 0.03 | 1.74 | 0.30 | m | 1.64 | 0.03 |
| 79 | Au | 0.82 | 0.04 | 1.01 | 0.18 | m | 0.82 | 0.04 |
| 80 | Hg | 1.19 | 0.08 | | | m | 1.19 | 0.08 |

**Table 3.** CI chondrite, solar (mainly photosphere), and recommended present-day solar abundances

|    |    | CI Chondrites | | Sun | | | Recommended | |
|----|----|------|------|------|------|------|------|------|
| Z  |    | A(X) |      | A(X) |      | Note | A(X) |      |
| 81 | Tl | 0.79 | 0.03 | 0.95 | 0.20 | m    | 0.79 | 0.03 |
| 82 | Pb | 2.06 | 0.03 | 2.00 | 0.06 | m    | 2.06 | 0.03 |
| 83 | Bi | 0.67 | 0.04 | 0.00 | 0.00 | m    | 0.67 | 0.04 |
| 90 | Th | 0.08 | 0.03 | 0.08 | UL   | m    | 0.08 | 0.03 |
| 92 | U  | -0.52| 0.03 | -0.47| UL   | m    | -0.52| 0.03 |

Abundances on the astronomical scale with log N(H)=12.
Note: a = average of meteoritic and solar value; m=meteoritic value; t = theoretical and/or indirectly determined.
UL: upper limit

Figure 4 shows the recommended abundances of the elements as a function of atomic number. The large abundances of H and He are not shown to avoid scale compression in the diagram. Overall, abundances decrease relatively smoothly with increasing atomic number. The regular pattern that odd numbered elements are less abundant than their even-numbered neighbors holds from the lightest to the heaviest elements; originally Harkins established this observation for the lighter elements up to the Fe region. Notable exceptions are the low abundances of Li, Be, and B, that consist of fragile nuclei that are easily destroyed in stellar interiors. The elemental abundance distribution is not controlled by the chemical properties of the elements but instead by nuclear properties.

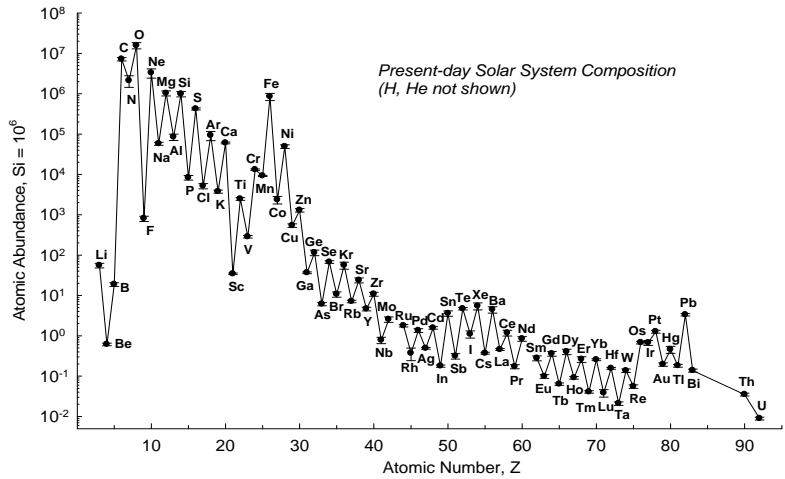

**Fig. 4.** Abundances of the elements as a function of atomic number

### 4.4 Mass Fractions X, Y, and Z in Present-Day Solar Material

Many applications in planetary sciences and astronomy use mass fractions of the elements rather than atomic abundances that we have dealt with so far. Mass fractions are also involved when the He abundance is to be derived. Although a value for He is listed in the tables above, the He abundance cannot be derived from the meteoritic nor the photospheric analyses. Thus, at this point one only has the atomic abundances of all elements relative to H or Si, except for He. Using the atomic weights of the elements, the relative atomic elemental abundances are converted to mass concentration ratios. Then the ratio of the mass sum of all heavy elements relative to the mass of H is obtained, which is needed to derive the He mass fraction from different constraints, and from that we finally obtain the atomic He abundance.

The mass fraction of H is usually abbreviated as X, that of He as Y, and the sum of the mass fractions of all other heavy elements as Z. The overall sum of these mass fractions is X+Y+Z=1. Absolute mass fractions of X, Y, and Z can be derived if the ratio of Z/X is known from atomic abundance analysis (the Z/X ratio can always be computed without knowing the He abundance), and if either the mass fraction of H or He is known independently.

The mass fraction of He can be inferred from inversion of helioseismic data by matching the sound speeds of H and He dominated mixtures under physical conditions appropriate for the solar convection zone. The mass fraction Z, which combines all other heavy elements is also important as it governs opacities and thus the density structure of the Sun's outer convection zone. The depth of the solar convection zone derived from helioseismic data poses constraints on the permissible fraction of heavy elements (a detailed review on helioseismology and the He abundance problem is given by Anita & Basu 2008).

The helioseismic inversion models require Z/X ratios and heavy element abundances (for opacities) as inputs. Therefore the He mass fraction and the He abundance from such models is not independent of X/Z. Ideally, one would use the abundance data and X/Z from the new compilation here to find the corresponding He abundance from helioseismic models and fits to solar data. One should not necessarily adopt a He abundance that is based on models that are calibrated to different Z/X than found for the new compilation of elemental abundances.

One can also obtain the absolute fractions of X, Y, and Z if the Z/X and the X (hydrogen mass fraction) are known. Basu & Anita (2004, 2008) have shown that the estimated mass faction of H from helioseismic models is relatively independent on Z/X ratios in the range of $0.0171 < Z/X < 0.0245$. Their models calibrated to Z/X=0.0171 and 0.0218 yield an average X=0.7389±0.0034 (Basu & Anita 2004). If the H mass fraction is indeed independent of the Z/X ratio, and if compositional variations *within* Z (mainly governed by the mass fractions of O, C, Ne, see below) also do not alter this conclusion much, we can use this X to estimate the He mass fraction.

With Z/X = 0.0191 found from the abundances in Table 3, and assuming X = 0.739, one obtains Z = 0.0141, and from this, Y = 1–X–Z = 0.2469. This mass fraction of He corresponds to an atomic He abundance of A(He) = 10.925.

**Table 4**. Present-day solar mass fractions and He abundance

| Present-Day | Z/X | X | Y | Z | A(He) |
|---|---|---|---|---|---|
| this work | 0.0191 | 0.7390 | 0.2469 | 0.0141 | 10.925 |
| A05, G07 | 0.0165 | *0.7383* | *0.2495* | 0.0122 | 10.93 |
| GS98 | 0.0231 | 0.7347 | 0.2483 | 0.0169 | 10.93 |

*Note*: AG89 cannot be done since AG89 do not give a present-day He abundance or Z/X. For protosolar values see below.

The mass faction of heavy elements (Z=0.014) is intermediate to those in the compilations by GS98 (Z=0.017) and G07 (Z=0.012); see Table 4 for a comparison for present-day solar values. The He mass fraction Y is smaller than that in previous compilations, but the (rounded) He abundance is the same for the three compilations in Table 4. The hydrogen mass fraction of X=0.739 (Basu & Anita 2004) adopted here is essentially the same as in A05/G07, and the smaller value for GS98 seems to be due to the different model assumptions for deriving the He abundance there.

**Table 5.** Concentration of present-day solar composition (mass %)

|  | this work | A05,G07 | GS98 |
|---|---|---|---|
| H (=X) | 73.90 | 73.92 | 73.47 |
| He (=Y) | 24.69 | 24.86 | 24.83 |
| O | 0.63 | 0.54 | 0.79 |
| C | 0.22 | 0.22 | 0.29 |
| Ne | 0.17 | 0.10 | 0.18 |
| Fe | 0.12 | 0.12 | 0.13 |
| N | 0.07 | 0.06 | 0.08 |
| Si | 0.07 | 0.07 | 0.07 |
| Mg | 0.06 | 0.06 | 0.07 |
| S | 0.03 | 0.03 | 0.05 |
| all other elements | 0.04 | 0.02 | 0.04 |
| total heavy elements (=Z) | 1.41 | 1.22 | 1.69 |

*Note*: Elements in order of decreasing concentration by mass.

Overall, the mass fractions derived here are closer to the ~10 year old GS98 compilation than to the most recent ones by L03, A05 or G07. Ideally, the He abundance proposed here needs to be evaluated with results from helioseismic models calibrated to the abundances of the elements (other than He) and the X/Z ratio found here.

The fraction for Z obtained here is higher than in the compilations by e.g., L03, A05, G07. The mass increase in Z should help to resurrect the standard solar models, which agreed with helioseimic constraints when the GS98 abundances were used, but crumbled under the too low Z values that were suggested in more recent compilations (see review by Basu & Anita 2008). Table 5 compares the mass fractions of the most abundant elements in present-day solar material (note that the order by mass is different from that by atomic abundances). About half of the mass fraction of Z is from O, followed by C, Ne, and Fe. The higher Z here comes mainly from increased O and Ne abundances, which may help to eradicate the problem with the incompatibility of standard solar models and recommended present-day solar abundances.

## 5 Solar System Abundances 4.56 Ga Ago

The data discussed above are for present-day abundances in the photosphere and meteorites. However, two processes affected the solar abundances over time. The first is element settling from the solar photosphere into the Sun's interior; the second is decay of radioactive isotopes that contribute to the overall atomic abundance of an element. The first, discussed in the following, is more important for the sun and large-scale modeling; the changes in isotopic compositions and their effects on abundances are comparably minor but important for radiometric dating. The isotopic effects are considered in the solar system abundance table in this section, but are not described at length here.

Settling or diffusion of heavy elements from the photosphere to the interior boundary layer of the convection zone and beyond lowered the elemental abundances (relative to H) from protosolar values 4.56 Ga ago (see Basu & Anita 2008). Over the Sun's lifetime, diffusion decreased abundances of elements heavier than He by ~13% from original protosolar values, whereas that of He dropped a little more by about ~15%; modeling these depletions also dependent on opacities, hence abundances. With these estimates, the proto-solar abundances (subscript 0) are calculated from the present-day data for the astronomical scales as

$$A(He)_0 = A(He) + 0.061,$$

and for all elements heavier than He it is

$$A(X)_0 = A(X) + 0.053 = \log N(X) + 1.586$$

The atomic abundances of the elements on both abundance scales are given in Table 6. Note that on the cosmochemical abundance scale ($N(Si)=10^6$), the relative abundances of the heavy elements do not change from the data in Table 3 because the scale is normalized to Si, one of the heavy elements. Only H and He change from present-day values: the relative H abundance is less, and the He abundance is lightly higher (because of the higher diffusive loss).

**Table 6.** Solar system abundances 4.56 Ga ago

| Z | | $N(Si)=10^6$ | | $\log N(H)=12$ | |
|---|---|---|---|---|---|
| 1 | H | 2.59E+10 | | 12.00 | |
| 2 | He | 2.51E+09 | 1.2E+08 | 10.986 | 0.02 |
| 3 | Li | 55.6 | 7.2 | 3.33 | 0.05 |
| 4 | Be | 0.612 | 0.043 | 1.37 | 0.03 |
| 5 | B | 18.8 | 1.9 | 2.86 | 0.04 |
| 6 | C | 7.19E+06 | 6.9E+05 | 8.44 | 0.04 |

**Table 6**. Solar system abundances 4.56 Ga ago

| Z  |    | N(Si)=$10^6$ |          | log N(H)=12 |      |
|----|----|----------|----------|----------|------|
| 7  | N  | 2.12E+06 | 6.8E+05  | 7.91 | 0.12 |
| 8  | O  | 1.57E+07 | 2.8E+06  | 8.78 | 0.07 |
| 9  | F  | 804      | 121      | 4.49 | 0.06 |
| 10 | Ne | 3.29E+06 | 8.5E+05  | 8.10 | 0.10 |
| 11 | Na | 57700    | 5100     | 6.35 | 0.04 |
| 12 | Mg | 1.03E+06 | 1.5E+05  | 7.60 | 0.06 |
| 13 | Al | 84600    | 15300    | 6.51 | 0.07 |
| 14 | Si | 1.00E+06 | 2E+04    | 7.59 | 0.08 |
| 15 | P  | 8300     | 1100     | 5.51 | 0.05 |
| 16 | S  | 4.21E+05 | 2.4E+04  | 7.21 | 0.02 |
| 17 | Cl | 5170     | 780      | 5.30 | 0.06 |
| 18 | Ar | 92700    | 24000    | 6.55 | 0.10 |
| 19 | K  | 3760     | 330      | 5.16 | 0.04 |
| 20 | Ca | 60400    | 3000     | 6.37 | 0.02 |
| 21 | Sc | 34.4     | 1.7      | 3.12 | 0.02 |
| 22 | Ti | 2470     | 200      | 4.98 | 0.03 |
| 23 | V  | 286      | 20       | 4.04 | 0.03 |
| 24 | Cr | 13100    | 500      | 5.70 | 0.02 |
| 25 | Mn | 9220     | 280      | 5.55 | 0.01 |
| 26 | Fe | 8.48E+05 | 1.69E+05 | 7.51 | 0.08 |
| 27 | Co | 2350     | 500      | 4.96 | 0.08 |
| 28 | Ni | 49000    | 5000     | 6.28 | 0.04 |
| 29 | Cu | 541      | 54       | 4.32 | 0.04 |
| 30 | Zn | 1300     | 130      | 4.70 | 0.04 |
| 31 | Ga | 36.6     | 1.8      | 3.15 | 0.02 |
| 32 | Ge | 115      | 18       | 3.65 | 0.06 |
| 33 | As | 6.10     | 0.55     | 2.37 | 0.04 |
| 34 | Se | 67.5     | 4.7      | 3.42 | 0.03 |
| 35 | Br | 10.7     | 1.6      | 2.62 | 0.06 |
| 36 | Kr | 55.8     | 11.3     | 3.33 | 0.08 |
| 37 | Rb | 7.23     | 0.51     | 2.45 | 0.03 |
| 38 | Sr | 23.3     | 1.6      | 2.95 | 0.03 |
| 39 | Y  | 4.63     | 0.50     | 2.25 | 0.04 |
| 40 | Zr | 10.8     | 1.2      | 2.62 | 0.04 |
| 41 | Nb | 0.780    | 0.139    | 1.48 | 0.07 |
| 42 | Mo | 2.55     | 0.40     | 1.99 | 0.06 |
| 44 | Ru | 1.78     | 0.11     | 1.84 | 0.03 |
| 45 | Rh | 0.370    | 0.128    | 1.15 | 0.13 |
| 46 | Pd | 1.36     | 0.15     | 1.72 | 0.04 |
| 47 | Ag | 0.489    | 0.024    | 1.28 | 0.02 |
| 48 | Cd | 1.57     | 0.11     | 1.78 | 0.03 |
| 49 | In | 0.178    | 0.012    | 0.84 | 0.03 |
| 50 | Sn | 3.60     | 0.54     | 2.14 | 0.06 |

**Table 6**. Solar system abundances 4.56 Ga ago

| Z | | N(Si)=10$^6$ | | log N(H)=12 | |
|---|---|---|---|---|---|
| 51 | Sb | 0.313 | 0.047 | 1.08 | 0.06 |
| 52 | Te | 4.69 | 0.33 | 2.26 | 0.03 |
| 53 | I | 1.10 | 0.22 | 1.63 | 0.08 |
| 54 | Xe | 5.46 | 1.10 | 2.32 | 0.08 |
| 55 | Cs | 0.371 | 0.019 | 1.16 | 0.02 |
| 56 | Ba | 4.47 | 0.81 | 2.24 | 0.07 |
| 57 | La | 0.457 | 0.023 | 1.25 | 0.02 |
| 58 | Ce | 1.18 | 0.19 | 1.66 | 0.06 |
| 59 | Pr | 0.172 | 0.020 | 0.82 | 0.05 |
| 60 | Nd | 0.856 | 0.043 | 1.52 | 0.02 |
| 62 | Sm | 0.267 | 0.013 | 1.01 | 0.02 |
| 63 | Eu | 0.10 | 0.01 | 0.58 | 0.04 |
| 64 | Gd | 0.360 | 0.049 | 1.14 | 0.06 |
| 65 | Tb | 0.06 | 0.00 | 0.39 | 0.03 |
| 66 | Dy | 0.404 | 0.062 | 1.19 | 0.06 |
| 67 | Ho | 0.09 | 0.01 | 0.55 | 0.03 |
| 68 | Er | 0.262 | 0.042 | 1.00 | 0.06 |
| 69 | Tm | 0.04 | 0.00 | 0.19 | 0.03 |
| 70 | Yb | 0.256 | 0.013 | 0.99 | 0.02 |
| 71 | Lu | 0.0380 | 0.0019 | 0.17 | 0.02 |
| 72 | Hf | 0.156 | 0.008 | 0.78 | 0.02 |
| 73 | Ta | 0.0210 | 0.0021 | -0.09 | 0.04 |
| 74 | W | 0.137 | 0.014 | 0.72 | 0.04 |
| 75 | Re | 0.0581 | 0.0058 | 0.35 | 0.04 |
| 76 | Os | 0.678 | 0.054 | 1.42 | 0.03 |
| 77 | Ir | 0.672 | 0.092 | 1.41 | 0.06 |
| 78 | Pt | 1.27 | 0.10 | 1.69 | 0.03 |
| 79 | Au | 0.195 | 0.019 | 0.88 | 0.04 |
| 80 | Hg | 0.458 | 0.092 | 1.25 | 0.08 |
| 81 | Tl | 0.182 | 0.015 | 0.85 | 0.03 |
| 82 | Pb | 3.31 | 0.23 | 2.11 | 0.03 |
| 83 | Bi | 0.138 | 0.012 | 0.73 | 0.04 |
| 90 | Th | 0.0440 | 0.0035 | 0.23 | 0.03 |
| 92 | U | 0.0238 | 0.0019 | -0.04 | 0.03 |

For completeness, the protosolar mass fractions X,Y, and Z are summarized in Table 7. However, as noted before for the present day solar mass fractions, it is up to the solar models and helioseismology to derive the best-fitting current and proto solar He mass fractions from the given abundances of the other elements.

**Table 7.** Protosolar mass fractions and He abundance

| | $Z_0/X_0$ | $X_0$ | $Y_0$ | $Z_0$ | $A(He)_0$ |
|---|---|---|---|---|---|
| this work [a] | 0.0215 | 0.7112 | 0.2735 | 0.0153 | 10.986 |
| A05, G07 [b] | 0.0185 | 0.7133 | 0.2735 | 0.0132 | 10.985 |
| GS98 [c] | 0.0231 | 0.7086 | 0.2750 | 0.0163 | 10.99 |
| AG89 [d] | 0.0267 | 0.7068 | 0.2743 | 0.0189 | 10.99 |

[a] present-day to protosolar values are converted using $(A(He)_0 = A(He) + 0.061$, all other elements (except H) from $A(X)_0 = A(X) + 0.053$

[b] G07 suggest $(A(He)_0 = A(He) + 0.057$, all other elements (except H) $A(X)_0 = A(X) + 0.05$.

[c] GS98: Changes in Z due to diffusion were not assumed ($Z/X = Z_0/X_0$); a ~10% loss of He from the photosphere was considered.

[d] Changes in Z due to diffusion were not assumed ($Z/X = Z_0/X_0$)

## 6 Abundances of the Nuclides

The abundance of an element is determined by the number and abundances of its stable isotopes, which in turn depends on the stability of the nuclei during thermonuclear reactions in stellar interiors. Already in the 1910s, Harkins and Oddo made the observation that elements with even atomic numbers are more abundant than their odd-numbered neighbors, which finds its explanation in the nuclear properties of the elements (see also Figure 4). An element is defined by its atomic number (Z), which is the number of positively charged nucleons (=protons) in its atoms. Atoms belonging to the same element may have different atomic masses due to a different number of neutral nucleons (=neutrons, neutron number N). In 1913, Soddy coined the term "isotope" for atoms with the same proton number but different neutron numbers after the Greek "isos-topos" meaning "at the same place" in the periodic table of the elements. The term isotope is in specific reference to a given element; whereas in a discussion of properties of atomic nuclei of different elements (with different Z and N) the generic term "nuclide" is usually more appropriate. However, often the terms "nuclide" and "isotope" are used as inter-exchangeable.

The sum of the number of protons (Z) and neutrons (N) is referred to as mass number A = Z+N. The mass number A is usually used when the nuclide abundance distributions are discussed, which is analogous to using the proton or atomic number Z for discussing elemental distributions.

There are 280 naturally occurring nuclides that make up the 83 stable and long-lived elements. These are all the elements up to Bi with Z=83, except for unstable Tc (Z=43) and Pm (Z=61) that only have short-lived isotopes, but the long-lived Th and U bring the total back to 83. Here "long-lived" or "short-lived" is with respect to the half-life of an isotope against radioactive decay and the age of the solar system. Long-lived means than an element is still present in measurable quantities since the solar system formed 4.6 Ga ago, and radioactive isotopes with half-lives above ~0.6 Ga usually qualify for this. Of the 280 nuclides, 266 are stable, and 14 have large half-lives such as $^{40}$K, $^{232}$Th, $^{235}$U, $^{238}$U, that find practical use in radiometric age dating of terrestrial and extraterrestrial samples, nuclear power reactors, and weaponry.

Considering only the atomic number, one finds that of the 83 elements, 43 have even Z, and 40 odd Z (note that Tc and Pm with only short-lived isotopes have odd atomic numbers, but Th and U have even ones), which reflects the higher stability of an atomic nucleus with even number of protons. This extends further to nuclei that also have an even number of neutrons. The proton and neutron numbers in the nuclei of the 266 stable nuclides lead to the following groupings:

Z even, N even: 159 nuclides
Z even, N odd: 53 nuclides
Z odd, N even: 50 nuclides
Z odd, N odd: 4 nuclides ($^{2}$H, $^{6}$Li, $^{10}$B, $^{14}$N)

Since an element's abundance is the sum of the abundances of the element's isotopes, a lower number of odd-Z numbered nuclides (50+4) than even-Z (159+53) means that there is a lower abundance of odd-Z elements. This is a simple explanation for the odd-even abundance distribution noted by Harkins and Oddo in the 1910s.

The mass numbers of the stable and long-lived nuclides range from A=1 ($^{1}$H) to 209 ($^{209}$Bi) except for gaps at A=5 and 8. After $^{209}$Bi we only have longer-lived nuclides of the actinides Th and U with the mass numbers 232, 235, and 238. Several nuclides have the same mass number but are isotopes of different elements, simply because A is given by Z+N. In comparisons of the nuclide distributions as function of mass number, the nuclides with the same A (isobaric nuclides, or isobars) are often summed up.

Table 8 summarizes the nuclide abundances and Figure 5 shows the abundance distribution of the nuclides as a function of mass number at the time of solar system formation 4.56 Ga ago. Figure 5 shows that nuclides with even A (shown as closed symbols) have usually higher abundances than the odd numbered nuclides (open symbols). Further, the odd numbered nuclides plot parallel to the even numbered A in a somewhat smoother distribution curve.

This behavior of nuclide distribution with mass number compares well to the behavior of elemental distribution with atomic number (Figure 4).

Abundances peak at mass numbers for closed proton and neutron "shells". These nuclear "shells" are analogous to the closed "electron shells" that characterize atomic properties. The "magic numbers" for nuclear stability are 2, 8, 20, 28, 50, 82, and 126; and nuclides with Z and/or N equal to these magic numbers are the ones that show large abundances in the diagram of abundance versus mass number (A=Z+N). This is particularly notable for the light doubly-magic nuclei with equal magic Z and N, e.g., $^{4}$He (Z=N=2), $^{16}$O(Z=N=8), and $^{40}$Ca (Z=N=20). Beyond the region of nuclides with mass numbers of 56 (the "Fe-peak" region), abundances decline more or less smoothly and spike at certain mass number regions. The nuclides beyond the Fe peak are products from neutron capture processes. The peaks in the distribution correspond to regions where either nuclides are preferentially made by the slow-neutron capture (S-) process operating in red giant stars (e.g, Y and Ba regions) or by the rapid-neutron capture (R-) process probably operating in supernovae (e.g., Pt region); see, e.g., Wallerstein et al. 1997, Woosley et al. 2002, Sneden et al. 2008 for reviews on stellar nucleosynthesis. Here the "slow" and "rapid" are in reference to beta-decay timescales of the intermediate, unstable nuclei produced during the neutron capture processes. The nuclide yields from these processes depend on the neutron energies and flux, but also on the abundance and stability of the target nuclei against neutron capture which in turn depends on Z and N. Hence the abundance distribution becomes controlled by the more stable "magic" nuclides that serve as bottlenecks for the overall yields in the neutron capture processes.

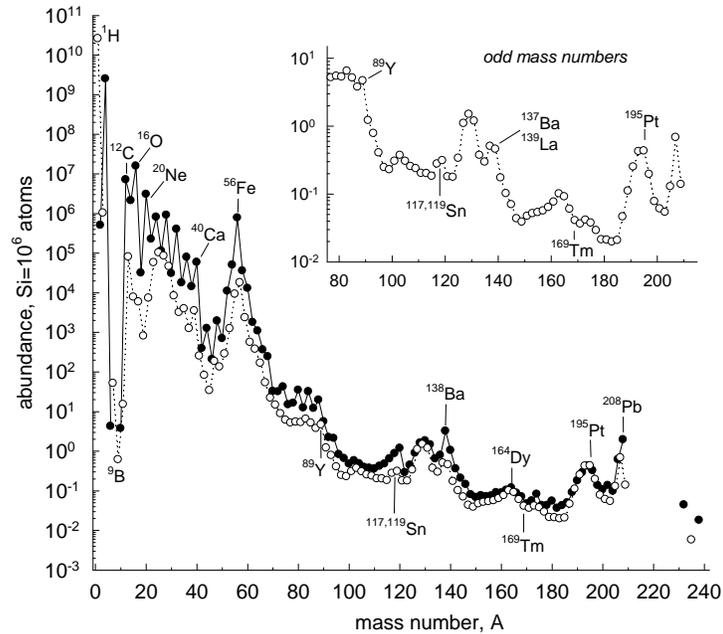

**Fig.5.** Solar system abundances of the nuclides 4.56 Ga ago (see LPG09).

Only a few nuclides beyond the Fe-group are exclusively produced by the S or R process; most nuclides have varying abundance contributions from both processes. If the contribution from each process for each isotope is known, the overall contribution of the R and S process to the elemental abundance can be estimated. The review on heavy element synthesis by Sneden 2008 includes a recent table on the R and S contributions to each element. A small number of proton-rich nuclides cannot be produced by the R and S process and are produced instead by the P-process, which probably involves neutrino induced disintegration of heavier nuclides. Like the R process, the P process is not yet completely understood. However, except for Mo, where P-process isotopes contribute about 25% to the elemental abundance, the contribution from P-process nuclides to overall elemental abundances is usually quite small.

Table 8 lists the percent contribution of the isotope(s) for each element, and the atomic abundance relative to $10^6$ silicon atoms at the time of solar system formation.

The abundances of radioactive isotopes (indicated by a star next to the element symbol) are adjusted accordingly. Table 8 is an update to the Table in L03, and includes several revisions of isotopic compositions, e.g., for Mo (Wieser & DeLaeter 2007), Dy (Segal et al. 2002, Chang et al. 2001), Er (Chang et al. 1998), Yb (DeLaeter & Bukilic 2006b), and Lu (DeLaeter & Bukilic 2006b).

**Table 8**. Solar system nuclide abundances 4.56 Gy ago

| Z | | A | atom% | N | Z | | A | atom% | N |
|---|---|---|---|---|---|---|---|---|---|
| 1 | H | 1 | 99.9981 | 2.59E+10 | 15 | P | 31 | 100 | 8300 |
| 1 | H | 2 | 0.00194 | 5.03E+05 | | | | | |
| | | | 100 | 2.59E+10 | 16 | S | 32 | 95.018 | 400258 |
| 2 | He | 3 | 0.0166 | 1.03E+06 | 16 | S | 33 | 0.75 | 3160 |
| 2 | He | 4 | 99.9834 | 2.51E+09 | 16 | S | 34 | 4.215 | 17800 |
| | | | 100 | 2.51E+09 | 16 | S | 36 | 0.017 | 72 |
| 3 | Li | 6 | 7.589 | 4.2 | | | | 100 | 421245 |
| 3 | Li | 7 | 92.411 | 51.4 | 17 | Cl | 35 | 75.771 | 3920 |
| | | | 100 | 55.6 | 17 | Cl | 37 | 24.229 | 1250 |
| 4 | Be | 9 | 100 | 0.612 | | | | 100 | 5170 |
| | | | | | 18 | Ar | 36 | 84.595 | 78400 |
| 5 | B | 10 | 19.820 | 3.7 | 18 | Ar | 38 | 15.381 | 14300 |
| 5 | B | 11 | 80.180 | 15.1 | 18 | Ar | 40 | 0.024 | 22 |
| | | | 100 | 18.8 | | | | 100 | 92700 |
| 6 | C | 12 | 98.889 | 7.11E+06 | 19 | K | 39 | 93.132 | 3500 |
| 6 | C | 13 | 1.111 | 7.99E+04 | 19 | K* | 40 | 0.147 | 6 |
| | | | 100 | 7.19E+06 | 19 | K | 41 | 6.721 | 253 |
| 7 | N | 14 | 99.634 | 2.12E+06 | | | | 100 | 3760 |
| 7 | N | 15 | 0.366 | 7.78E+03 | 20 | Ca | 40 | 96.941 | 58500 |
| | | | 100 | 2.12E+06 | 20 | Ca | 42 | 0.647 | 391 |
| 8 | O | 16 | 99.763 | 1.57E+07 | 20 | Ca | 43 | 0.135 | 82 |
| 8 | O | 17 | 0.037 | 5.90E+03 | 20 | Ca | 44 | 2.086 | 1260 |
| 8 | O | 18 | 0.200 | 3.15E+04 | 20 | Ca | 46 | 0.004 | 2 |
| | | | 100 | 1.57E+07 | 20 | Ca | 48 | 0.187 | 113 |
| 9 | F | 19 | 100 | 804 | | | | 100 | 60400 |
| | | | | | 21 | Sc | 45 | 100 | 34.4 |
| 10 | Ne | 20 | 92.9431 | 3.06E+06 | | | | | |
| 10 | Ne | 21 | 0.2228 | 7.33E+03 | 22 | Ti | 46 | 8.249 | 204 |
| 10 | Ne | 22 | 6.8341 | 2.25E+05 | 22 | Ti | 47 | 7.437 | 184 |
| | | | 100 | 3.29E+06 | 22 | Ti | 48 | 73.72 | 1820 |
| 11 | Na | 23 | 100 | 57700 | 22 | Ti | 49 | 5.409 | 134 |
| | | | | | 22 | Ti | 50 | 5.185 | 128 |
| 12 | Mg | 24 | 78.992 | 8.10E+05 | | | | 100 | 2470 |
| 12 | Mg | 25 | 10.003 | 1.03E+05 | 23 | V | 50 | 0.2497 | 0.7 |
| 12 | Mg | 26 | 11.005 | 1.13E+05 | 23 | V | 51 | 99.7503 | 285.7 |
| | | | 100 | 1.03E+06 | | | | 100 | 286.4 |
| 13 | Al | 27 | 100 | 8.46E+04 | 24 | Cr | 50 | 4.3452 | 569 |
| 14 | Si | 28 | 92.230 | 9.22E+05 | 24 | Cr | 52 | 83.7895 | 11000 |
| 14 | Si | 29 | 4.683 | 4.68E+04 | 24 | Cr | 53 | 9.5006 | 1240 |
| 14 | Si | 30 | 3.087 | 3.09E+04 | 24 | Cr | 54 | 2.3647 | 309 |
| | | | 100 | 1.00E+06 | | | | 100 | 13100 |

| Z | | A | atom% | N | Z | | A | atom% | N |
|---|---|---|---|---|---|---|---|---|---|
| 25 | Mn | 55 | 100 | 9220 | 35 | Br | 79 | 50.686 | 5.43 |
| | | | | | 35 | Br | 81 | 49.314 | 5.28 |
| 26 | Fe | 54 | 5.845 | 49600 | | | | 100 | 10.7 |
| 26 | Fe | 56 | 91.754 | 7.78E+05 | 36 | Kr | 78 | 0.362 | 0.20 |
| 26 | Fe | 57 | 2.1191 | 18000 | 36 | Kr | 80 | 2.326 | 1.30 |
| 26 | Fe | 58 | 0.2819 | 2390 | 36 | Kr | 82 | 11.655 | 6.51 |
| | | | 100 | 8.48E+05 | 36 | Kr | 83 | 11.546 | 6.45 |
| 27 | Co | 59 | 100 | 2350 | 36 | Kr | 84 | 56.903 | 31.78 |
| | | | | | 36 | Kr | 86 | 17.208 | 9.61 |
| 28 | Ni | 58 | 68.0769 | 33400 | | | | | |
| 28 | Ni | 60 | 26.2231 | 12900 | | | | 100 | 55.8 |
| 28 | Ni | 61 | 1.1399 | 559 | 37 | Rb | 85 | 70.844 | 5.121 |
| 28 | Ni | 62 | 3.6345 | 1780 | 37 | Rb* | 87 | 29.156 | 2.108 |
| 28 | Ni | 64 | 0.9256 | 454 | | | | 100 | 7.23 |
| | | | 100 | 49000 | 38 | Sr | 84 | 0.5580 | 0.13 |
| 29 | Cu | 63 | 69.174 | 374 | 38 | Sr | 86 | 9.8678 | 2.30 |
| 29 | Cu | 65 | 30.826 | 167 | 38 | Sr | 87 | 6.8961 | 1.60 |
| | | | 100 | 541 | 38 | Sr | 88 | 82.6781 | 19.2 |
| 30 | Zn | 64 | 48.63 | 630 | | | | 100 | 23.3 |
| 30 | Zn | 66 | 27.9 | 362 | 39 | Y | 89 | 100 | 4.63 |
| 30 | Zn | 67 | 4.1 | 53 | | | | | |
| 30 | Zn | 68 | 18.75 | 243 | 40 | Zr | 90 | 51.452 | 5.546 |
| 30 | Zn | 70 | 0.62 | 8 | 40 | Zr | 91 | 11.223 | 1.210 |
| | | | 100 | 1300 | 40 | Zr | 92 | 17.146 | 1.848 |
| 31 | Ga | 69 | 60.108 | 22.0 | 40 | Zr | 94 | 17.38 | 1.873 |
| 31 | Ga | 71 | 39.892 | 14.6 | 40 | Zr | 96 | 2.799 | 0.302 |
| | | | 100 | 36.6 | | | | 100 | 10.78 |
| 32 | Ge | 70 | 21.234 | 24.3 | 41 | Nb | 93 | 100 | 0.780 |
| 32 | Ge | 72 | 27.662 | 31.7 | 42 | Mo | 92 | 14.525 | 0.370 |
| 32 | Ge | 73 | 7.717 | 8.8 | 42 | Mo | 94 | 9.151 | 0.233 |
| 32 | Ge | 74 | 35.943 | 41.2 | 42 | Mo | 95 | 15.838 | 0.404 |
| 32 | Ge | 76 | 7.444 | 8.5 | 42 | Mo | 96 | 16.672 | 0.425 |
| | | | 100 | 115 | 42 | Mo | 97 | 9.599 | 0.245 |
| 33 | As | 75 | 100 | 6.10 | 42 | Mo | 98 | 24.391 | 0.622 |
| | | | | | 42 | Mo | 100 | 9.824 | 0.250 |
| 34 | Se | 74 | 0.89 | 0.60 | | | | 100 | 2.55 |
| 34 | Se | 76 | 9.37 | 6.32 | 44 | Ru | 96 | 5.542 | 0.099 |
| 34 | Se | 77 | 7.64 | 5.15 | 44 | Ru | 98 | 1.869 | 0.033 |
| 34 | Se | 78 | 23.77 | 16.04 | 44 | Ru | 99 | 12.758 | 0.227 |
| 34 | Se | 80 | 49.61 | 33.48 | 44 | Ru | 100 | 12.599 | 0.224 |
| 34 | Se | 82 | 8.73 | 5.89 | 44 | Ru | 101 | 17.060 | 0.304 |

| Z | | A | atom% | N | Z | | A | atom% | N |
|---|---|---|---|---|---|---|---|---|---|
| 44 | Ru | 102 | 31.552 | 0.562 | 52 | Te | 120 | 0.096 | 0.005 |
| 44 | Ru | 104 | 18.621 | 0.332 | 52 | Te | 122 | 2.603 | 0.122 |
| | | | 100 | 1.78 | 52 | Te | 123 | 0.908 | 0.043 |
| 45 | Rh | 103 | 100 | 0.370 | 52 | Te | 124 | 4.816 | 0.226 |
| | | | | | 52 | Te | 125 | 7.139 | 0.335 |
| 46 | Pd | 102 | 1.02 | 0.0139 | 52 | Te | 126 | 18.952 | 0.889 |
| 46 | Pd | 104 | 11.14 | 0.1513 | 52 | Te | 128 | 31.687 | 1.486 |
| 46 | Pd | 105 | 22.33 | 0.3032 | 52 | Te | 130 | 33.799 | 1.585 |
| 46 | Pd | 106 | 27.33 | 0.371 | | | | 100 | 4.69 |
| 46 | Pd | 108 | 26.46 | 0.359 | 53 | I | 127 | 100 | 1.10 |
| 46 | Pd | 110 | 11.72 | 0.159 | 54 | Xe | 124 | 0.129 | 0.007 |
| | | | 100 | 1.36 | 54 | Xe | 126 | 0.112 | 0.006 |
| 47 | Ag | 107 | 51.839 | 0.254 | 54 | Xe | 128 | 2.23 | 0.122 |
| 47 | Ag | 109 | 48.161 | 0.236 | 54 | Xe | 129 | 27.46 | 1.499 |
| | | | 100 | 0.489 | 54 | Xe | 130 | 4.38 | 0.239 |
| 48 | Cd | 106 | 1.25 | 0.020 | 54 | Xe | 131 | 21.80 | 1.190 |
| 48 | Cd | 108 | 0.89 | 0.014 | 54 | Xe | 132 | 26.36 | 1.438 |
| 48 | Cd | 110 | 12.49 | 0.197 | 54 | Xe | 134 | 9.66 | 0.527 |
| 48 | Cd | 111 | 12.8 | 0.201 | 54 | Xe | 136 | 7.87 | 0.429 |
| 48 | Cd | 112 | 24.13 | 0.380 | | | | 100 | 5.46 |
| 48 | Cd | 113 | 12.22 | 0.192 | 55 | Cs | 133 | 100 | 0.371 |
| 48 | Cd | 114 | 28.73 | 0.452 | | | | | |
| 48 | Cd | 116 | 7.49 | 0.118 | 56 | Ba | 130 | 0.106 | 0.005 |
| | | | 100 | 1.57 | 56 | Ba | 132 | 0.101 | 0.005 |
| 49 | In | 113 | 4.288 | 0.008 | 56 | Ba | 134 | 2.417 | 0.108 |
| 49 | In | 115 | 95.712 | 0.170 | 56 | Ba | 135 | 6.592 | 0.295 |
| | | | 100 | 0.178 | 56 | Ba | 136 | 7.853 | 0.351 |
| 50 | Sn | 112 | 0.971 | 0.035 | 56 | Ba | 137 | 11.232 | 0.502 |
| 50 | Sn | 114 | 0.659 | 0.024 | 56 | Ba | 138 | 71.699 | 3.205 |
| 50 | Sn | 115 | 0.339 | 0.012 | | | | 100 | 4.471 |
| 50 | Sn | 116 | 14.536 | 0.524 | 57 | La* | 138 | 0.091 | 0.000 |
| 50 | Sn | 117 | 7.676 | 0.277 | 57 | La | 139 | 99.909 | 0.457 |
| 50 | Sn | 118 | 24.223 | 0.873 | | | | 100 | 0.457 |
| 50 | Sn | 119 | 8.585 | 0.309 | 58 | Ce | 136 | 0.186 | 0.002 |
| 50 | Sn | 120 | 32.593 | 1.175 | 58 | Ce | 138 | 0.250 | 0.003 |
| 50 | Sn | 122 | 4.629 | 0.167 | 58 | Ce | 140 | 88.450 | 1.043 |
| 50 | Sn | 124 | 5.789 | 0.209 | 58 | Ce | 142 | 11.114 | 0.131 |
| | | | 100 | 3.60 | | | | 100 | 1.180 |
| 51 | Sb | 121 | 57.213 | 0.179 | 59 | Pr | 141 | 100 | 0.172 |
| 51 | Sb | 123 | 42.787 | 0.134 | | | | | |
| | | | 100 | 0.313 | | | | | |

| Z | | A | atom% | N | Z | | A | atom% | N |
|---|---|---|---|---|---|---|---|---|---|
| 60 | Nd | 142 | 27.044 | 0.231 | 68 | Er | 166 | 33.503 | 0.088 |
| 60 | Nd | 143 | 12.023 | 0.103 | 68 | Er | 167 | 22.869 | 0.060 |
| 60 | Nd | 144 | 23.729 | 0.203 | 68 | Er | 168 | 26.978 | 0.071 |
| 60 | Nd | 145 | 8.763 | 0.075 | 68 | Er | 170 | 14.910 | 0.039 |
| 60 | Nd | 146 | 17.130 | 0.147 | | | | 100 | 0.262 |
| 60 | Nd | 148 | 5.716 | 0.049 | 69 | Tm | 169 | 100 | 0.0406 |
| 60 | Nd | 150 | 5.596 | 0.048 | | | | | |
| | | | 100 | 0.856 | 70 | Yb | 168 | 0.12 | 0.0003 |
| 62 | Sm | 144 | 3.073 | 0.008 | 70 | Yb | 170 | 2.98 | 0.0076 |
| 62 | Sm* | 147 | 14.993 | 0.041 | 70 | Yb | 171 | 14.09 | 0.0361 |
| 62 | Sm* | 148 | 11.241 | 0.030 | 70 | Yb | 172 | 21.69 | 0.0556 |
| 62 | Sm | 149 | 13.819 | 0.037 | 70 | Yb | 173 | 16.10 | 0.0413 |
| 62 | Sm | 150 | 7.380 | 0.020 | 70 | Yb | 174 | 32.03 | 0.0821 |
| 62 | Sm | 152 | 26.742 | 0.071 | 70 | Yb | 176 | 13.00 | 0.0333 |
| 62 | Sm | 154 | 22.752 | 0.060 | | | | 100 | 0.256 |
| | | | 100 | 0.267 | 71 | Lu | 175 | 97.1795 | 0.0370 |
| 63 | Eu | 151 | 47.81 | 0.0471 | 71 | Lu* | 176 | 2.8205 | 0.0011 |
| 63 | Eu | 153 | 52.19 | 0.0514 | | | | 100 | 0.0380 |
| | | | 100 | 0.0984 | 72 | Hf | 174 | 0.162 | 0.0003 |
| 64 | Gd | 152 | 0.203 | 0.0007 | 72 | Hf | 176 | 5.206 | 0.0081 |
| 64 | Gd | 154 | 2.181 | 0.0078 | 72 | Hf | 177 | 18.606 | 0.0290 |
| 64 | Gd | 155 | 14.800 | 0.0533 | 72 | Hf | 178 | 27.297 | 0.0425 |
| 64 | Gd | 156 | 20.466 | 0.0736 | 72 | Hf | 179 | 13.629 | 0.0212 |
| 64 | Gd | 157 | 15.652 | 0.0563 | 72 | Hf | 180 | 35.100 | 0.0547 |
| 64 | Gd | 158 | 24.835 | 0.0894 | | | | 100 | 0.156 |
| 64 | Gd | 160 | 21.864 | 0.0787 | 73 | Ta | 180 | 0.0123 | 2.6E-06 |
| | | | 100 | 0.360 | 73 | Ta | 181 | 99.9877 | 0.0210 |
| 65 | Tb | 159 | 100 | 0.0634 | | | | 100 | 0.0210 |
| | | | | | 74 | W | 180 | 0.120 | 0.0002 |
| 66 | Dy | 156 | 0.056 | 0.0002 | 74 | W | 182 | 26.499 | 0.0363 |
| 66 | Dy | 158 | 0.095 | 0.0004 | 74 | W | 183 | 14.314 | 0.0196 |
| 66 | Dy | 160 | 2.329 | 0.0094 | 74 | W | 184 | 30.642 | 0.0420 |
| 66 | Dy | 161 | 18.889 | 0.0762 | 74 | W | 186 | 28.426 | 0.0390 |
| 66 | Dy | 162 | 25.475 | 0.1028 | | | | 100 | 0.137 |
| 66 | Dy | 163 | 24.896 | 0.1005 | 75 | Re | 185 | 35.662 | 0.0207 |
| 66 | Dy | 164 | 28.260 | 0.1141 | 75 | Re* | 187 | 64.338 | 0.0374 |
| | | | 100 | 0.404 | | | | 100 | 0.0581 |
| 67 | Ho | 165 | 100 | 0.0910 | 76 | Os | 184 | 0.020 | 0.0001 |
| | | | | | 76 | Os | 186 | 1.598 | 0.0108 |
| 68 | Er | 162 | 0.139 | 0.0004 | 76 | Os | 187 | 1.271 | 0.0086 |
| 68 | Er | 164 | 1.601 | 0.0042 | 76 | Os | 188 | 13.337 | 0.0904 |

| Z | | A | atom% | N |
|---|---|---|---|---|
| 76 | Os | 189 | 16.261 | 0.110 |
| 76 | Os | 190 | 26.444 | 0.179 |
| 76 | Os | 192 | 41.070 | 0.278 |
| | | | 100 | 0.678 |
| 77 | Ir | 191 | 37.272 | 0.250 |
| 77 | Ir | 193 | 62.728 | 0.421 |
| | | | 100 | 0.672 |
| 78 | Pt* | 190 | 0.014 | 0.0002 |
| 78 | Pt | 192 | 0.783 | 0.010 |
| 78 | Pt | 194 | 32.967 | 0.420 |
| 78 | Pt | 195 | 33.832 | 0.431 |
| 78 | Pt | 196 | 25.242 | 0.322 |
| 78 | Pt | 198 | 7.163 | 0.091 |
| | | | 100 | 1.27 |
| 79 | Au | 197 | 100 | 0.195 |
| 80 | Hg | 196 | 0.15 | 0.001 |
| 80 | Hg | 198 | 9.97 | 0.046 |
| 80 | Hg | 199 | 16.87 | 0.077 |
| 80 | Hg | 200 | 23.10 | 0.106 |
| 80 | Hg | 201 | 13.18 | 0.060 |
| 80 | Hg | 202 | 29.86 | 0.137 |
| 80 | Hg | 204 | 6.87 | 0.031 |
| | | | 100 | 0.458 |
| 81 | Tl | 203 | 29.524 | 0.054 |
| 81 | Tl | 205 | 70.476 | 0.129 |
| | | | 100 | 0.182 |
| 82 | Pb | 204 | 1.997 | 0.066 |
| 82 | Pb | 206 | 18.582 | 0.614 |
| 82 | Pb | 207 | 20.563 | 0.680 |
| 82 | Pb | 208 | 58.858 | 1.946 |
| | | | 100 | 3.306 |
| 83 | Bi | 209 | 100 | 0.1382 |
| 90 | Th* | 232 | 100 | 0.0440 |
| 92 | U* | 234 | 0.002 | 4.9E-07 |
| 92 | U* | 235 | 24.286 | 0.0058 |
| 92 | U* | 238 | 75.712 | 0.0180 |
| | | | 100 | 0.0238 |

**Acknowledgements**: I thank Bruce Fegley and Herbert Palme for discussions and comments.